**Title**
Neuronal coupling by endogenous electric fields: Cable theory and applications to coincidence detector neurons in the auditory brainstem

**Authors**
Joshua H. Goldwyn[1,2,3]
John Rinzel[1,2]

**Contributions**
JHG and JR designed study, analyzed data, and wrote manuscript. JHG performed simulations.

**Affiliations**
[1] Center for Neural Science, New York University, New York, NY, United States
[2] Courant Institute of Mathematical Sciences, New York University, New York, NY, United States
[3] Department of Mathematics, Ohio State University, Columbus, OH, United States

**Running Head**
Neuronal coupling by endogenous electric fields

**Address for Correspondence**
Joshua H. Goldwyn
100 Math Tower
231 West 18th Avenue
Columbus OH, 43210-1174
Phone: 614-292-5256
Fax: 614 292-1479
Email: jhgoldwyn@gmail.com

**Abstract**

The ongoing activity of neurons generates a spatially- and time-varying field of extracellular voltage ($V_e$). This $V_e$ field reflects population-level neural activity, but does it modulate neural dynamics and the function of neural circuits? We provide a cable theory framework to study how a bundle of model neurons generates $V_e$ and how this $V_e$ feeds back and influences membrane potential ($V_m$). We find that these "ephaptic interactions" are small but not negligible. The model neural population can generate $V_e$ with millivolt-scale amplitude and this $V_e$ perturbs the $V_m$ of "nearby" cables and effectively increases their electrotonic length. After using passive cable theory to systematically study ephaptic coupling, we explore a test case: the medial superior olive (MSO) in the auditory brainstem. The MSO is a possible locus of ephaptic interactions: sounds evoke large $V_e$ *in vivo* in this nucleus (millivolt-scale). The $V_e$ response is thought to be generated by MSO neurons that perform a known neuronal computation with submillisecond temporal precision (coincidence detection to encode sound source location). Using a biophysically-based model of MSO neurons, we find millivolt-scale ephaptic interactions consistent with the passive cable theory results. These subtle membrane potential perturbations induce changes in spike initiation threshold, spike time synchrony, and time difference sensitivity. These results suggest that ephaptic coupling may influence MSO function.

**Keywords**
Ephaptic coupling, medial superior olive, cable theory, extracellular voltage, coincidence detection

# INTRODUCTION

Neurons are bathed in a shared extracellular voltage ($V_e$) that is generated by voltage-gated ionic currents, synaptic currents, and other transmembrane currents (see Buzsáki et al. 2012 for review). This endogenous $V_e$ can be recorded *in vivo* using depth electrodes (e.g. the local field potential) and is detectable on the surface of the scalp (e.g. electroencephalography). It has long been known that applied (i.e. "exogenous") $V_e$, delivered via a stimulating electrode, can evoke and modulate neural activity (Strumwasser and Rosenthal 1960, e.g.). Indeed, this is the basic mechanism by which neural prostheses such as cochlear implants and deep brain stimulation provide therapeutic benefits. The functional consequences of endogenous $V_e$, however, remain a subject of investigation. In this work, we present an idealized model for assessing neuronal coupling via endogenous $V_e$. We develop a cable theoretic framework that shows how millivolt-scale endogenous $V_e$ can induce millivolt-scale perturbations in membrane potential. We apply these insights to the medial superior olive in the auditory brainstem and find (in simulations) that ephaptic coupling can alter spike activity in these specialized coincidence detector neurons.

Following Arvanitaki (1942), we refer to interactions between neurons mediated by endogenous extracellular voltage as *ephaptic* interactions. She had "no doubt that the activity of an element in the midst of a cell agglomeration can influence that of its neighbors [via ephaptic coupling]" (Arvanitaki 1942). Classical studies of interactions between side-by-side axons supported her assertion (Katz and Schmitt 1940; Arvanitaki 1942; Ramón and Moore 1978) and recent studies have demonstrated that relatively weak and oscillatory applied $V_e$ (designed to be "endogenous-like") can enhance spike time synchrony *in vitro* (Radman et al. 2007; Frölich and McCormick 2010; Anastassiou et al. 2011).

In this work, we present a systematic study of ephaptic coupling in a bundle of dendrites using methods that are an extension of classical cable theory (Rall 1977). By invoking assumptions of cable theory and by considering a population of identical neurons (see Materials and Methods), we solve the ephaptic coupling problem by computing intra- and extracellular voltages in coupled one-dimensional domains. This approach differs from standard "line-source" or "point-source" approximations in which $V_e$ is computed from simulated neural activity without including any feedback between $V_e$ and neural activity (Klee and Rall 1977; Holt and Koch 1999; Lindén et al. 2011; Reimann et al. 2013). Our method is substantially simplified compared to solvers that couple membrane dynamics of three-dimensional neuron models to Maxwell's equations for extracellular electric and magnetic fields (Malik 2011; Agudelo-Toro and Neef 2013).

We consider a population of identical model neurons that receive identical inputs and that are arranged with spatial symmetry (i.e. equally spaced and oriented parallel to one another). These assumptions appear restrictive, but the approach is inspired by pioneering analyses of endogenous $V_e$ in the olfactory bulb (Rall and Shepherd 1968) and cerebellum (Nicholson and Llinás 1971). We have found this idealized modeling approach useful to describe sound-evoked extracellular voltages recorded *in vivo* in the auditory brainstem (Goldwyn et al. 2014).

The auditory brainstem $V_e$ is an intriguing test case for studying ephaptic interactions. Neurons in the medial superior olive (MSO) are believed to generate the auditory brainstem $V_e$ (Galambos et al. 1959; Biedenbach and Freeman 1964; Moushegian et al. 1964; Tsuchitani and Boudreau 1964; Clark and Dunlop 1968; Mc Laughlin et al. 2010; Goldwyn et al. 2014). The dense packing of MSO neurons' dendrites and the presence of a prominent, sound-evoked extracellular voltage field have led to suggestions that ephaptic interactions may be at work in this nucleus (Schwartz 1977; Ashida and Carr 2011). MSO neurons perform a known computation: sound localization via temporally precise coincidence detection of dendritic inputs (see Grothe et al. 2010 for review). We can evaluate the functional consequences of ephaptic interactions in this system by simulating coincidence detection in MSO neuron models in the presence of (simulated) endogenous $V_e$.

MSO neurons are bipolar (two dendrites with modest branching, extending away from a central soma, Rautenberg et al. 2009). Back-propagating action potentials in the soma are small (Scott et al. 2007; Franken et al. 2015) and are difficult to detect with extracellular and juxtacellular electrodes (Yin and Chan 1990; van der Heijden et al. 2013). It is reasonable, therefore, to begin our study with a passive cable model and neglect contributions of spike-generating Na currents. In this initial analysis we gain insights into ephaptic interactions in the MSO and dendritic bundles in general. We find that the cable population can generate millivolt-scale $V_e$ and that this $V_e$ can induce a millivolt-scale perturbation in the membrane potential of a neuron embedded in the $V_e$ bath. These ephaptic interactions are largest for electrotonically compact cables (i.e. large cable space constant).

The passive cable results illustrate that $V_e$, due to its spatially-distributed nature, can hyperpolarize or depolarize different portions of a "nearby" neuron. As a corollary, when we extend our MSO neuron model to include spike-generating Na currents, we find that ephaptic coupling can have either "excitatory" or "inhibitory" effects depending on the location of spike initiation in the spatially-varying $V_e$. The relatively modest (millivolt-scale) perturbations of membrane potential due to endogenous $V_e$ alter spike initiation threshold, spike timing, and the sensitivity of MSO neurons to arrival times of bilateral inputs. Specifically, we find that ephaptic coupling suppresses spiking activity if spikes are generated near the soma of the MSO neuron model and promotes spiking activity if spikes are generated at locations distant from the soma. Our results establish, in principle, that ephaptic coupling can influence neural processing in this early stage of the auditory pathway.

## Glossary

*General*
| | |
|---|---|
| $t$ | Time [ms] |
| $x$ | Distance [cm] |
| $V_i$ | Intracellular voltage [mV] |
| $V_e$ | Extracellular voltage [mV] |
| $V_m = V_i - V_e$ | Transmembrane potential [mV] |
| $E_{lk}$ | Reversal potential of leak current [mV] |
| $r_i$ | Intracellular (core) resistance per unit length [Ω/cm] |
| $r_e$ | Extracellular resistance across per unit length [Ω/cm] |
| $N$ | Number of neurons in population |
| $\kappa = Nr_e/r_i$ | Coupling constant |
| $\rho = R_e/R_i$ | Resistivity ratio |
| $\delta$ | Packing density |
| $A_i, A_e$ | Cross-sectional area of intra- and extracellular domains [cm$^2$] |
| $\wedge$ | Parameter or variable associated with "test" neuron ($\hat{V}_m$, e.g.) |

*Passive cable model*

| | |
|---|---|
| $c_m$ | Capacity per unit length [mF/cm] |
| $r_m$ | Resistance across a unit length [Ωcm] |
| $\tau = c_m r_m$ | Membrane time constant [ms] |
| $\lambda = \sqrt{r_m/r_i}$ | Cable space constant [cm] |
| $\lambda_{ref}$ | "Reference value" of cable space constant [cm] |
| $T = t/\tau$ | Nondimensional time variable |
| $X = x/\lambda$ | Nondimensional space variable |
| $L = l/\lambda$ | Nondimensional length of cable |
| $i_{in}$ | Input current per unit length [mA/cm] |
| $i_m$ | Membrane current per unit length [mA/cm] |
| $l$ | Physical length of cable [cm] |
| $d_g$ | Distance from end of cable to electric ground [cm] |
| $\omega$ | Frequency of sinusoidal current input (Hz) |
| $U_m, U_i, U_e$ | Voltage in Fourier domain |

*Medial superior olive model*

| | |
|---|---|
| $C_m$ | Capacitance [mF/cm$^2$] |
| $R_i$ | Intracellular (axial) resistivity [Ωcm] |
| $R_e$ | Extracellular resistivity [Ωcm] |
| $I_i, I_e, I_m, I_{in}$ | Intracellular, extracellular, transmembrane, and input current [mA] |
| $J_{in}$ | Input current density [mA/cm$^2$] |
| $G_{lk}, G_{KLT}, G_h, G_{Na}$ | Maximal conductance density for leak, KLT, h, and Na currents [mS/cm$^2$] |
| $E_K, E_h, E_{Na}$ | Reversal potential for K, h and Na currents [mV] |
| $w, z$ | Gating variables for KLT current |
| $m, h$ | Gating variables for Na current |
| $u_\infty$ | Steady-state functions for gating variables [where $u = w, z, m$, or $h$] |
| $\tau_u$ | Time constants for activation and inactivation variables [where $u = w, z, m$, or $h$] |
| $G_{syn}$ | Maximal synaptic conductance density [mS/cm$^2$] |
| $E_{syn}$ | Reversal potential of synaptic current [mV] |
| $\tau_{syn}$ | Synaptic conductance time constant [ms] |
| $t_0$ | Onset time of synaptic event [ms] |
| $g_{axial}$ | Axial conductance between soma and spike initiation zone [nS] |
| $d$ | Diameter of MSO neuron model [cm] |
| $\Delta x$ | Length of compartment in discretized MSO neuron model [cm] |
| $S$ | Surface area of compartment [cm$^2$] |

# MATERIALS AND METHODS
## Ephaptic coupling in a population of passive cables

*Model Formulation:* In the standard (passive) cable theory, spatio-temporal dynamics of membrane voltage $V_m(x,t)$ are governed by the balance of capacitative, leak, and applied currents crossing the cell membrane and diffusion of current within the cell (Rall 1977):

$$c_m \frac{\partial V_m}{\partial t} = -\frac{1}{r_m}(V_m - E_{lk}) + \frac{1}{r_i}\frac{\partial^2 V_i}{\partial x^2} + i_{in}. \tag{1}$$

$c_m$ is the membrane capacity per unit length [mF/cm], $r_m$ is the resistance across a unit length of membrane [Ωcm], $r_i$ is intracellular (core) resistance per unit length [Ω/cm], $E_{lk}$ is the leak current reversal potential [mV], and $i_{in}$ is input current per unit length [mA/cm]. At the ends of the cable, we impose a "sealed-end" (zero axial current) boundary condition by requiring $\partial V_i/\partial x = 0$ for $x = 0$ or $l$ where $l$ is the physical length of the cable. In all cable model simulations, we will present $V_m$ as its deviation from resting potential (or, equivalently, set $E_{lk} = 0$ mV).

To gauge the effect of $V_e$ on $V_m$, one can substitute $V_m + V_e$ for $V_i$ in Eq. 1:

$$c_m \frac{\partial V_m}{\partial t} = -\frac{1}{r_m}(V_m - E_{lk}) + \frac{1}{r_i}\frac{\partial^2 V_m}{\partial x^2} + \frac{1}{r_i}\frac{\partial^2 V_e}{\partial x^2} + i_{in}. \tag{2}$$

In this formulation, the effect of $V_e$ on the dynamics of $V_m$ appears as a spatially distributed current source proportional to the second spatial derivative of $V_e$. Positive curvature of $V_e$ acts locally as a depolarizing ("excitatory") current. This is the basis for the *activating function* method (Rattay 1986), a heuristic used to approximate the effect of applied extracellular fields on neurons, for instance in studies of neural prostheses (Rattay 1999). Holt and Koch (1999) refer to the second spatial derivative of $V_e$ divided by $r_i$ as a "fictitious distributed current (the ephaptic current)."

A common modeling assumption is that extracellular voltage has a negligible impact on the cell's voltage dynamics. In that case, since $V_m = V_i - V_e$, one sets $V_e = 0$ for all $x$ (no "ephaptic current") and lets $V_i = V_m$. We are specifically investigating how $V_e$ affects neuronal dynamics via ephaptic interactions so we retain $V_i$ in Eq. 1 as a quantity distinct from $V_m$.

We now describe how the activity of neurons generates $V_e$ and how this $V_e$ feeds back and influences $V_m$ in an idealized model of $N$ identical and parallel cables. If all such cables receive similar input $i_{in}$ (in terms of temporal dynamics and spatial location), then the spatio-temporal distribution of membrane currents will be similar across the population. As a consequence, $V_e$ in the region surrounding any one cable will be similar to $V_e$ surrounding any neighboring cable in the population. In other words, the gradient of $V_e$ (which is proportional to the current flow in the extracellular region) will be directed, for the most part, parallel to the orientation of the cables. We thus make the assumption that the extracellular space can be described as one-dimensional volume conductor. This reduces the problem of modeling extracellular interactions to two, coupled one-dimensional domains: the inside of the cable (intracellular core conductor) and a thin layer surrounding the cable (extracellular volume conductor).

In the one-dimensional extracellular region, there is a current balance relationship comprised of the sum of all membrane currents in the population of $N$ cables $i_m = \sum_{n=1}^{N}\left[c_m \frac{\partial V_m^{(n)}}{\partial t} + \frac{1}{r_m}\left(V_m^{(n)} - E_{lk}\right) - i_{in}^{(n)}\right]$ and current flow along the one-dimensional extracellular pathway $-\frac{1}{r_e}\frac{\partial^2 V_e}{\partial x^2}$, where $r_e$ is the resistance per unit length [$\Omega$/cm] in the extracellular region in the direction parallel to the cables. The superscript $(n)$ is the index of neurons in the population. Under the assumption described above, $i_{in}^{(n)}$ (and, therefore, $V_m^{(n)}$) are similar for all $N$ cables, so we divide by $N$ and obtain the population-averaged current balance relation

$$c_m \frac{\partial V_m}{\partial t} = -\frac{1}{r_m}(V_m - E_{lk}) - \frac{1}{N r_e}\frac{\partial^2 V_e}{\partial x^2} + i_{in}. \quad (3)$$

$V_m$ and $i_{in}$ represent population-averaged quantities, for instance $V_m = \frac{1}{N}\sum_{n=1}^{N} V_m^{(n)}$. Note that the intracellular current balance relation in Eq. 1 still holds if $V_m$, $V_i$, and $i_{in}$ are population-averaged quantities, so going forward we will maintain this mean-field perspective. A more general formulation would allow the applied current $i_{in}$ to differ in the intracellular and extracellular domains (cf. Tuckwell 1988). We consider $i_{in}$ to be transmembrane current (synaptic current, for example) so we require the applied intra- and extracellular currents to be identical.

Extracellular space extends beyond the ends of the cables and volume conduction allows $V_e$ to spread to a distant electric ground at which $V_e = 0$ mV. We impose a mixed boundary condition that describes the flow of current to electric ground along a one-dimensional current pathway of length $d_g$. These boundary conditions are $d_g \partial V_e/\partial x - V_e = 0$ at $x = 0$ and $d_g \partial V_e/\partial x + V_e = 0$ at $x = l$.

Using a standard re-parameterization, we define the time and space constants of the cable (Rall 1977): $\tau = c_m r_m$ and $\lambda^2 = r_m/r_i$. In addition, we introduce a coupling parameter $\kappa = N r_e/r_i$. The governing equations for the coupled, intracellular/extracellular system are:

$$\tau \frac{\partial V_m}{\partial t} = -(V_m - E_{lk}) + \lambda^2 \frac{\partial^2 V_i}{\partial x^2} + r_m i_{in}. \quad (4)$$

$$\tau \frac{\partial V_m}{\partial t} = -(V_m - E_{lk}) - \frac{\lambda^2}{\kappa} \frac{\partial^2 V_e}{\partial x^2} + r_m i_{in}. \tag{5}$$

$$\left. \frac{\partial V_i(x,t)}{\partial x} \right|_{x=0,l} = 0 \tag{6}$$

$$d_e \frac{\partial V_e(0,t)}{\partial x} - V_e(0,t) = 0 \text{ and } d_e \frac{\partial V_e(l,t)}{\partial x} + V_e(l,t) = 0 \tag{7}$$

We will often report results in terms of the nondimensional spatial variable $X = x/\lambda$ and the cable length $L = x/\lambda$.

To further investigate ephaptic coupling, we embed an additional neuron, with possibly different cable properties and input current, into the surrounding $V_e$. We ignore its $O(1/N)$ contribution to $V_e$ and the population-averaged $V_m$, but since membrane potential is the difference between intracellular and extracellular voltage, $V_e$ perturbs this additional neuron's membrane potential. We refer to this as a "test neuron" and its membrane and intracellular voltages satisfy Eq. 4 and the boundary condition in Eq. 6. We use the ^ accent to indicate parameter values for the test neuron that differ from the cable population. We note that there is no spike generating mechanism in the passive cable model. We view this as a subthreshold model and are neglecting any contributions of spiking activity to $V_e$ and ephaptic interactions.

*Solution Method:* Equations 4-7 form a system of partial differential algebraic equations (PDAEs) and, in general, require special solution methods (Lucht et al 1997a, 1997b). For simple cases (constant or sinusoidal input current $i_{in}$ injected at a single point on the cable), we reformulate and solve these equations numerically as a boundary value problem. For general current waveforms (and voltage-gated membrane currents), numerical solution methods are available to integrate these equations in time (see solution method for MSO model, below).

In response to a constant current input, the system will reach a steady state with $\partial V_m/\partial t = 0$. This eliminates the time dependence in Eqs. 4 and 5. Steady state spatial profiles of $V_i$ and $V_m$ satisfy a linear, constant coefficient system of ordinary differential equations:

$$\lambda^2 V_i'' = V_i - V_e - E_{lk} - r_m i_{in} \tag{8}$$

$$\frac{\lambda^2}{\kappa} V_e'' = -(V_i - V_e - E_{lk}) + r_m i_{in} \tag{9}$$

$$\hat{\lambda}^2 \hat{V}_i'' = \hat{V}_i - V_e - \hat{E}_{lk} - \hat{r}_m \hat{i}_{in} \tag{10}$$

with boundary conditions given in Eqs. 6 and 7. Derivatives in these equations are with respect to the spatial variable $x$ and the ^ accent indicates parameters and variables associated with the test neuron. We solve this boundary value problem using the function `bvp4c` in MATLAB (R2012b, The MathWorks Inc.).

We can also formulate a boundary value problem to describe the frequency-response characteristics of the coupled intracellular-extracellular system. In response to the stimulus $i_{in} = i_0 \exp(i2\pi\omega t) \delta(x-x_0)$, intracellular and extracellular voltages are of the form $V_i(x) = U_i(x)\exp(i2\pi\omega t)+E_{lk}$ and $V_e(x) = U_e(x)\exp(i2\pi\omega t)$ and solve the following ordinary differential equations:

$$\lambda^2 U_i'' = (1 + i2\pi\tau\omega)(U_i - U_e) - r_m i_0 \delta(x - x_0) \tag{11}$$

$$\frac{\lambda^2}{\kappa} U_e'' = -(1 + i2\pi\tau\omega)(U_i - U_e) + r_m i_0 \delta(x - x_0) \tag{12}$$

$$\lambda^2 U_i'' = (1 + i2\pi\tau\omega)(U_i - U_e) - r_m i_0 \delta(x - x_0). \tag{13}$$

The boundary conditions in Eqs. 6 and 7 are applied to the amplitude variables $U(x)$ and we solve these equations using `bvp4c` in MATLAB. We report the amplitude of the oscillatory response to $i_{in}$ as the absolute value of $U(x)$ and use the `phase` command in MATLAB to recover the phase of the response. Note that the intracellular amplitude $U_i$ measures the deviation of the intracellular voltage from the resting potential $E_{lk}$. In all simulations, we will present cable model $V_m$ responses in terms of their deviation from rest (equivalent to setting $E_{lk}$ = 0 mV in above equations).

**Ephaptic coupling in an idealized model of the medial superior olive**

*Model formulation:* Neurons of the MSO are thought to generate prominent extracellular voltages in response to acoustic stimuli (Galambos et al. 1959, e.g.). In previous work, we developed an MSO model that predicted spatiotemporal features of these extracellular voltage responses (Goldwyn et al. 2014). Here, we adapt this model to test ephaptic interactions among MSO neurons. The MSO neuron model differs from the passive cable discussed above because it includes voltage-gated membrane current in addition to the leak current and a non-uniform morphology (two dendrites, connected to a soma). These nonlinearities and inhomogeneities interfere with the construction of a population-averaged model.

However, we have argued previously that a mean-field perspective is justified for describing *in vivo* MSO responses to pure tone stimuli (Goldwyn et al. 2014). Specifically, we noted that MSO neurons have a relatively simple morphology (bipolar dendrites with minimal branching) and are oriented roughly in parallel. Moreover, early stages of the auditory pathway are specialized to deliver inputs to MSO neurons with high levels of phase locking (Joris et al. 1994, e.g.). We idealize these anatomical and physiological observations to argue that MSO neurons are arranged with spatial symmetry and their inputs arrive in synchrony with one another across a local subpopulation. These conditions of synchrony and symmetry justify the use of a simplified one-dimensional volume conductor model (Rall and Shepherd 1968).

The current balance relations for the intra- and extracellular domains of the MSO model (in terms of current density) are:

$$C_m \frac{\partial V_m}{\partial t} = G_{lk}(V_m - E_{lk}) + G_{KLT}w^4z(V_m - E_K) + G_h(V_m - E_h) + \frac{d}{4R_i}\frac{\partial^2 V_i}{\partial x^2} + J_{in} \quad (14)$$

$$C_m \frac{\partial V_m}{\partial t} = G_{lk}(V_m - E_{lk}) + G_{KLT}w^4z(V_m - E_K) + G_h(V_m - E_h) + \frac{d}{\kappa 4R_i}\frac{\partial^2 V_e}{\partial x^2} + J_{in} \quad (15)$$

with the same boundary conditions as the passive model (Eqs. 6 and 7). These equations and the dynamics of the gating variables $w$ and $z$ are adapted from a biophysically-based model of an MSO neuron first presented by Mathews et al. (2010) (see Eqs. 16-19 below). We have used the coupling coefficient $\kappa$ in Eq. 15 as we did in Eq. 5 of the passive cable model to avoid the introduction of unknown parameters for extracellular resistance and population size. We will discuss plausible values for $\kappa$ (see Fig. 1).

The neuron model consists of two dendrites extending away from a central soma. Each dendrite is a 150 μm-long cylinder with diameter $d$ = 3.5 μm. The soma is a cylinder of length 20 μm and diameter $d$ = 20 μm. Membrane capacitance is $C_m$ = 0.9 μF/cm$^2$, intracellular (axial) resistivity is $R_i$ = 200 Ωcm, and leak conductance density is $G_{lk}$ = 0.3 mS/cm$^2$. The neuron model includes low threshold K (KLT) current and hyperpolarization-activated cation (h) current. Maximal KLT conductance density is 17 mS/cm$^2$ in the soma and 3.58 mS/cm$^2$ in the dendrites, and maximal h conductance density is 0.86 mS/cm$^2$ in the soma and 0.18 mS/cm$^2$ in the dendrites. These parameters correspond to the "step-gradient" model in Mathews et al. (2010).

The KLT current has a voltage-gated activation variable $w$ and inactivation variable $z$ that evolve according to $du/dt = [u_\infty(V_m)-u]/\tau_u(V_m)$ where $u = w$ or $z$ and

$$w_\infty(V_m) = \left[1 + e^{-(V_m+57.34)/11.7}\right]^{-1} \quad (16)$$

$$\tau_w(V_m) = 21.5\left[6e^{(V_m+60)/7} + 24e^{-(V_m+60)/50.6}\right]^{-1} + 0.35 \quad (17)$$

$$z_\infty(V_m) = 0.73\left[1 + e^{(V_m+67)/6.16}\right]^{-1} + 0.27 \quad (18)$$

$$\tau_z(V_m) = 170\left[5e^{(V_m+60)/10} + e^{-(V_m+70)/8}\right]^{-1} + 10.7. \quad (19)$$

The gating variable for the $h$-current evolves slowly (time scale on the order of hundreds of milliseconds, see Khurana et al. 2011). We make the simplification, therefore, that $G_h$ remains at a constant value in all simulations.

In most simulations the input current $J_{in}$ is a simulated synaptic input with alpha function conductance:

$$I_{in} = G_{syn}\left(\frac{t-t_0}{\tau_{syn}}\right)\exp\left(1-\frac{t-t_0}{\tau_{syn}}\right)(V_m - E_{syn}). \tag{20}$$

$G_{syn}$ is the maximal synaptic conductance density [mS/cm$^2$], $t_0$ is the onset time of the synaptic event, $\tau_{syn}$ is the synaptic time constant, and $E_{syn}$ is the reversal potential. Excitatory inputs to the MSO neurons are fast (Golding and Oertel 2012) and primarily target dendrites (Couchman et al. 2012). We set the synaptic time constant in Eq. 20 to $\tau_{syn}$ = 0.2 ms, consistent with *in vitro* physiology and previous modeling studies (Jercog et al. 2010; Fischl et al. 2012; Myoga et al. 2014) and the reversal potential is 0 mV. In some simulations the onset times $t_0$ of synaptic events are fixed (with a specific timing difference between inputs on the two dendrites, for instance) and in other simulations the onset times are drawn from Poisson processes to approximate more realistic input patterns. Exceptions are the simulations of spike time synchrony and time-difference tuning curves (Fig. 11) in which the alpha function conductance in Eq. 20 is replaced with a rectified sine wave that is meant to approximate the population-averaged conductance input to MSO in response to pure tone stimuli. We place excitatory inputs on either dendrite ~125 μm from the soma. MSO neurons also receive inhibitory inputs (Grothe and Sanes 1993, 1994) that primarily target the soma (Couchman et al. 2012). We omit these in the current study. We have explored the contribution of inhibition to simulated $V_e$ responses in an MSO model in previous work (Goldwyn et al. 2014).

To highlight ephaptic coupling in the MSO model we embed an additional "test" neuron in $V_e$. This single neuron's contribution to $V_e$ is $O(1/N)$ so can be neglected. We keep the properties of this test neuron the same as those in the population, but in some simulations we attach an additional compartment to the soma that contains spike-generating Na current. We use this test neuron to evaluate how ephaptic interactions alter coincidence detection in the MSO neuron model.

The additional compartment represents the putative spike initiation zone (SIZ). This likely corresponds to the axon initial segment and/or a proximal node of Ranvier (see Lehnert et al. (2014) for a recent computational study of spike initiation in an MSO neuron model). The membrane dynamics in the SIZ are

$$C_m \frac{\partial V_m^{SIZ}}{\partial t} = G_{LK}(V_m^{SIZ} - E_{LK}) + G_{Na}m^3h(V_m - E_{Na}) + \frac{g_{axial}}{S_{SIZ}}(V_i^{SIZ} - V_i^{soma}). \tag{21}$$

$V_i^{SIZ}$ is the membrane potential in the SIZ, $V_m^{SIZ}$ is the SIZ membrane potential, and $V_i^{soma}$ is the intracellular potential at the soma. The SIZ is influenced by the local extracellular voltage because $V_m^{SIZ} = V_i^{SIZ} - V_e$, where $V_e$ is the extracellular voltage at the location of the SIZ.

The dynamics of the Na current (*m* and *h* variables) are modified from the Rothman and Manis model (2003). They are adjusted for a temperature of 35°C (Khurana et al. 2011) and the gating properties of the Na inactivation variable *h* are "left-shifted" by 6 mV. This modification is motivated by *in vitro* measurements (Scott et al. 2010) and enhances the phasic character of the model neuron (Huguet et al. 2012).

$$m_\infty(V_m) = \left[1 + e^{-(V_m+38)/7}\right]^{-1} \tag{22}$$

$$\tau_m(V_m) = 0.24\left[\frac{10}{5e^{(V_m+60)/18} + 36e^{-(V_m+60)/25}} + 0.04\right] \tag{23}$$

$$h_\infty(V_m) = \left[1 + e^{(V_m+71)/6}\right]^{-1} \tag{24}$$

$$\tau_h(V_m) = 0.24\left[\frac{100}{7e^{(V_m-60)/11} + 10e^{-(V_m+60)/25}} + 0.6\right] \tag{25}$$

The SIZ is assumed to be a small patch of membrane (1 μm diameter and 1 μm length, surface area is $S_{siz}$ = 3.14 μm²) with a dense concentration of Na channels. The leak conductance density is $G_{lk}$ = 200 mS/cm², the maximum Na conductance density is $G_{Na}$ = 75,000 mS/cm². The capacitance per area is $C_m$ = 0.9 μF/cm². These parameters are chosen so that the characteristics of the backpropagating action potential in the soma are similar to *in vitro* recordings (Scott et al. 2007). The current balance equation for the soma of the test neuron is also altered to account for axial current flow to and from the SIZ with $g_{axial}$ = 60 nS. Reasonable assumptions are that the initial segment's diameter is ~1-1.5 μm (Lehnert et al. 2014) and the axial resistance connecting the SIZ to the soma is 200 Ωcm. For these values and $g_{axial}$ = 60 nS, the implied length of the soma-to-SIZ connection is ~6-15 μm, consistent with the anatomy of the initial segment reported by Lehnert et al. In our simulations we take a phenomenological view of the SIZ and allow its location in the $V_e$ field to be more distant from the soma. The intracellular connection between the SIZ and the soma (i.e. $g_{axial}$) remains the same in all simulations regardless of the location of the SIZ in the extracellular domain.

We assume that spikes do not contribute significantly to $V_e$. We do not, therefore, include spikes in the MSO population and do not take into account spikes generated in the SIZ of the test neuron when computing $V_e$. This assumption is based on a consensus that post-synaptic membrane currents in the MSO generate the prominent, ongoing sound-evoked $V_e$ in the auditory brain and spikes do not significantly contribute to it (Galambos et al. 1959, Mc Laughlin et al. 2010, e.g.). Physiological observations support the assumption that spikes do not contribute to $V_e$. Back-propagating action potentials in the soma of MSO neuron are small (~20 mV) when measured *in vitro* (Scott et al. 2007) and *in vivo* (Franken et al. 2015) and can be difficult to detect in extracellular recordings (Yin and Chan 1990) and juxtacellular recordings (van der Heijden et al 2013).

*Solution Method:* As mentioned above, Eqs. 14 and 15 represent a system of PDAEs. Due to the voltage–gated ion currents and the synaptic (conductance) input, the coupling between intra- and extracellular voltages is nonlinear and $V_m$-dependent. To solve these equations, we discretize the spatial domain in small bins of length $\Delta x$. This converts the PDAEs into a system of differential algebraic equations (DAEs) that can be solved with appropriate software (we use SUNDIALS, available at http://computation.llnl.gov/casc/sundials/main.html, Hindmarsh et al. 2005). The discretization is analogous to the compartmental method for computing $V_m$ dynamics in a spatially-extended neuron model (Segev and Burke 1998), but in our formulation intracellular and extracellular compartments reside at each point in discretized space and are coupled to one another. The numerical method is designed to conserve the flow of current in and out of each compartment, so it is necessary to define intra-, extra-, and membrane currents ($I_i$, $I_e$, $I_m$) [units of milliamperes].

Let $x$ identify the spatial location of a compartment of width $\Delta x$. Then we denote the intracellular (axial) current flow from an adjacent compartment into the compartment at $x$ as:

$$I_i(x - \Delta x/2, t) = \frac{V_i(x - \Delta x, t) - V_i(x, t)}{r_i \Delta x} \tag{26}$$

and extracellular current flow as:

$$I_e(x - \Delta x/2, t) = \frac{V_e(x - \Delta x, t) - V_e(x, t)}{\kappa r_i \Delta x}. \tag{27}$$

For notational simplicity we have substituted $r_i$ for $4R_i/\pi d^2$ in these and subsequent equations. Note that, unlike the cable equation, the diameter $d$ of each compartment is not uniform (soma is larger than dendrites). As a result, $r_i$ and the coupling parameter $\kappa$ are larger in the soma compartments than in the dendrite compartments. For ease of notation we do not explicitly indicate this $x$-dependence of $r_i$ and $\kappa$.

The net flows of intra- and extracellular currents at location $x$ satisfy current balance relations with the transmembrane current $I_m(x,t)$:

$$I_m(x, t) = \frac{V_i(x - \Delta x, t) - 2V_i(x, t) + V_i(x + \Delta x, t)}{r_i \Delta x} \tag{28}$$

$$I_m(x, t) = \frac{V_e(x - \Delta x, t) - 2V_e(x, t) + V_e(x + \Delta x, t)}{\kappa r_i \Delta x}. \tag{29}$$

The transmembrane current $I_m(x,t)$ at location $x$ (by convention, outward flow of positive ions is positive current) consists of capacitative, leak, ionic, and input currents:

$$I_m(x,t) = S\left[C_m \frac{dV_m}{dt} - g_{lk}(V_m - E_{lk}) + g_{Na} m^3 h (V_m - E_{Na}) + g_K r (V_m - E_K)\right] + I_{in} \tag{30}$$

$S$ denotes surface area of the MSO compartment and it depends on $x$; it is larger in soma compartments than in dendrite compartments. $I_{in}$ is the input current in units of milliamperes.

By identifying the right side of Eq. 30 with the second-order differences in Eqs. 28 and 29, we obtain a set of equations that dynamically couple intracellular and extracellular voltage via the membrane potential. We impose boundary conditions as in the cable model: a "sealed end" for $V_i$ in the intracellular domain and a linear decay of $V_e$ to ground (0 mV) at a distance of 1 mm from the ends of the dendrites. To include the test neuron with SIZ, we solve two versions of the intracellular model, one for the population MSO response (the generators of $V_e$) and one for the test neuron. The test neuron includes coupling to its SIZ given by Eq. 21. The SUNDIALS numerical solver steps forward in time while maintaining these current balance relations by using a variable order, variable coefficient implicit method (Hindmarsh et al. 2005). We used a relative tolerance of $10^{-6}$ and an absolute error tolerance of $10^{-8}$ in the solver and obtained the solution at 1μs time steps.

MATLAB code is available on the ModelDB repository and includes example solutions of the MSO model and user-friendly simulation code for the passive cable model.

# RESULTS
### Ephaptic interactions in passive cables

*Remarks on coupling parameter κ:* The coupling parameter $\kappa = Nr_e/r_i$ dictates the strength of interactions between $V_e$ and $V_m$. The standard cable theory assumes that $V_e$ is spatially uniform and negligible. This is the limiting case of $\kappa \to 0$. To estimate a range of plausible non-zero values of $\kappa$, we introduce two new parameters: the packing density $\delta$ of the population of $N$ cables and the ratio of extracellular to intracellular volume resistivities $\rho = R_e/R_i$. Let $A_i$ be the cross-sectional area of each cable and $A_e$ the cross-sectional area of the extracellular space (i.e. the total cross-sectional area of the brain region under consideration is $NA_i+A_e$). Then, the relations $r_u = R_u/A_u$ ($u = i, e$) and $\delta = NA_i/(NA_i+A_e)$ allow us to express the coupling parameter as $\kappa = \rho\delta/(1-\delta)$. The advantage of this formulation is that we can estimate bounds on $\rho$ and $\delta$.

Extracellular resistivity $R_e$ has been measured in a number of biological tissues and animals (Geddes and Baker 1967, e.g.). Based on these and other experiments, modeling studies typically use values of $\rho$ that range from ~1 to 4. In simulations of neuron-$V_e$ interactions, Holt and Koch (1999) used $\rho = 2.2$ ($R_i = 150$ Ωcm and $R_e = 330$ Ωcm). Recent modeling studies of cortical local field potentials have used similar values: $\rho = 2.2$ in Lindén et al. (2010) and $\rho = 3.5$ in Reimann et al. (2013), for example.

The packing density $\delta$ of a local population of neurons depends on their spatial arrangement and morphology. For the idealized case of uniform cables oriented in parallel to one another we can provide a theoretical upper bound by treating the population of cables as an example of circle packing in a plane (when viewed in cross-section). In this case, the theoretical upper bound for δ is ~0.9 (Weisstein). If the extracellular domain is a circle then the upper bound for δ would be closer to 0.8 for a moderate (<100) number of cables (Graham et al., 1998). Figure 1 shows a contour plot of κ as a function of $\rho$ and $\delta$ in these parameter ranges. For small packing densities and small $\rho$ the coupling strength approaches 0, but $\kappa$ exceeds 1 in over half of the parameter space and can reach values as large as ~15. Unless otherwise specified, we will gauge ephaptic effects in the cable model by comparing simulations without $V_e$ coupling ($\kappa = 0$) to simulations with $\kappa = 1$. Our estimates for $\kappa$ in the MSO model are also within this plausible range. They are marked in Fig. 1 and discussed below.

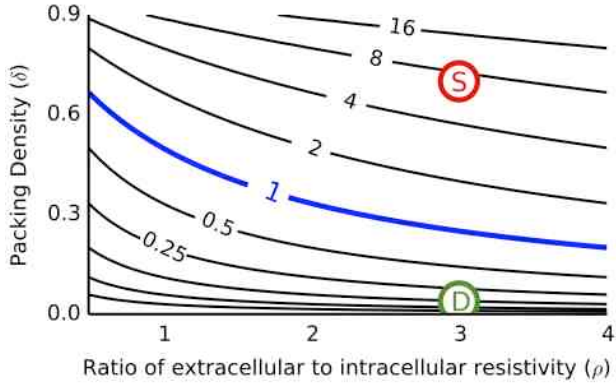

**Figure 1:** Contour plot of the coupling parameter $\kappa$ for plausible values of packing density ($\delta$) and resistance ratio ($\rho$). Coupling parameter is $\kappa = \rho\delta/(1-\delta)$, contour lines have logarithmic spacing (powers of 2). We use $\kappa = 1$ (blue contour) in simulations of the cable model with ephaptic coupling. Estimates for $\kappa$ in the MSO model are marked with colored circles. $\kappa$ is larger surrounding the soma (7, red S) than the dendrite (0.12, green D) since the larger diameter of the soma implies a larger packing density ($\delta = 0.7$ for soma and 0.038 for dendrites; $\rho = 3$ throughout).

*Responses to constant current – Initial observations:* We begin our investigation of ephaptic interactions by studying responses to constant current applied at a single point along the cable. Responses in this scenario are solutions to the boundary value problem for the coupled intra- and extracellular voltages at steady state given in Eqs. 8-10. The space constant in MSO neurons is roughly the length of one dendrite (Mathews et al. 2010). We use the cable to gain initial insights that can be applied to the MSO, so we set the space constant $\lambda$ in the cable model to be one-half the physical length of the cable, unless otherwise indicated. Steady state solutions do not depend on the cable time constant $\tau$.

We first injected a constant, depolarizing current to a site near the left end of the cables in the population (schematic in Fig. 2A). The population-averaged $V_m$ is maximal at the site of the input and attenuates with distance along the cable (Fig. 2B). The stimulus amplitude in these simulations is arbitrary; response amplitude scales linearly with stimulus amplitude for current injection to passive cables.

The effect of coupling all cables in the population via $V_e$ is evident in the difference between the population-averaged $V_m$ response in the absence of coupling ($\kappa = 0$, black line in Fig. 2B) and the response with ephaptic coupling ($\kappa = 1$, blue line in Fig. 2B). Ephaptic coupling increases the membrane depolarization $V_m$ near the stimulation site, acting to increase the input resistance. At locations distant from the input site, ephaptic coupling decreases the depolarization of $V_m$. The net result of ephaptic coupling is to increase the rate at which $V_m$ attenuates with distance along the cable. In other words, ephaptic coupling decreases the cable space constant, effectively increasing the electrotonic length of the cable. This effect of $V_m$-to-$V_e$ coupling was noted by Rall in his example of an "axon in oil" (Rall 1977). In that analysis of an infinite cable surrounded by a thin extracellular layer, Rall noted that the space constant of the cable decreases as extracellular resistance increases according to $\sqrt{r_m/(r_i + r_e)}$, or equivalently $\sqrt{(R_m/\pi d)/(R_i/A_i + R_e/A_e)}$ where $d$ is the cable diameter.

We highlight the ephaptic effect by showing the response of a "test neuron" embedded in the $V_e$ field generated by all neurons in the population. The test neuron's membrane potential $V_m$ (more precisely, the deviation of $V_m$ from rest) displays the same changes discussed above: a local depolarization near the site of input current and hyperpolarization at more distant locations on the cable (Fig. 2C). This cable has identical properties to the cables whose population-averaged $V_m$ is shown in Fig. 2B. The test neuron does not receive any current injection. The $V_e$ surrounding the test neuron is the sole "input" that determines the spatial profile of $V_m$ in Fig. 2C. We point out that ephaptic coupling hyperpolarizes $V_m$ at the center of the cable ($V_m$ < 0 mV at $x/\lambda$ = 1). This anticipates a main finding in our MSO simulations: $V_e$ produced by dendritic excitation can have a hyperpolarizing or "inhibitory" effect on the soma of a "nearby" MSO neuron.

We can understand the ephaptic effects in these simulations by examining the spatial profile of $V_e$ in Fig. 2D. The input current is a transfer of positive ions from the extracellular domain into the interior of the population-averaged cable and thus $V_e$ is negative near input site; the input acts as a current sink to the surrounding extracellular domain. Recall from Eq. 2 the heuristic that the second spatial derivative of $V_e$ acts as a distributed "ephaptic current" to perturb membrane potential (neglecting boundary effects). The second spatial derivative of $V_e$ at this minimum is positive and thus the test neuron "feels" $V_e$ as a depolarizing current at the input location. Conservation of current in the system requires that the flow of current from the extracellular domain into the cables (i.e. the stimulus current) must be returned back to the extracellular domain at other locations along the cable. As a result, $V_e$ is positive at spatial locations distant from the stimulus site. At these distant locations the extracellular space draws current out of the test neuron and hyperpolarizes $V_m$; the efflux acts as a current source to the extracellular surround.

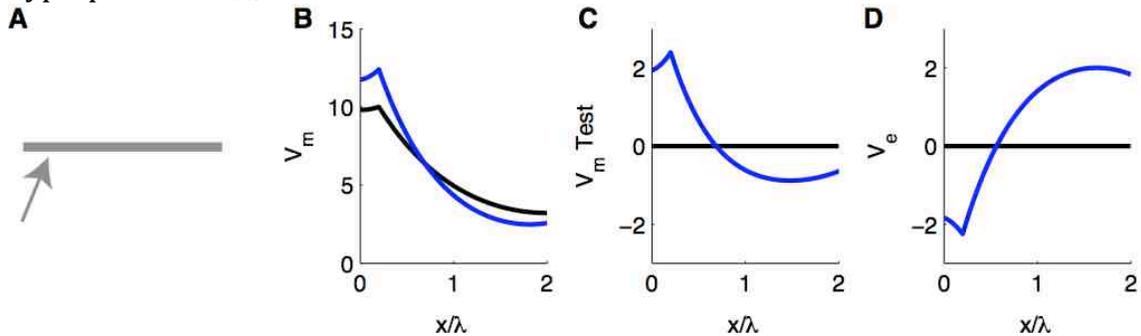

**Figure 2:** Passive cable response to constant input current. Responses without ephaptic coupling (κ = 0) shown in black, responses with ephaptic coupling (κ = 1) shown in blue. **(A)** Input location is near one end of cable. **(B)** Steady-state depolarization of population-averaged $V_m$. **(C)** Steady-state $V_m$ response of "test neuron" that is coupled to population of cables via the extracellular voltage. $V_m$ in B and C are plotted as deviations from resting potential. **(D)** Steady-state $V_e$ response produced by population of cables. Abscissa in all panels is distance along the cable normalized by the cable space constant $\lambda$ (one-half the physical length of the cable).

In these simulations, a relatively modest input current and coupling parameter (~12 mV maximum depolarization in population-averaged response, κ =

1) produces $V_e$ amplitudes of ±2 mV. The amplitude of the response of the test neuron is similar to $V_e$: ~2.5 mV maximum depolarization and ~-1 mV hyperpolarization. This result is consistent with the observation made by Anastassiou et al. (2010) that the spatial frequency of $V_e$ must be sufficiently large, relative to the electric and physical lengths of the cable, to perturb $V_m$. More precisely, they showed that a spatially inhomogeneous $V_e$ with spatial frequency $f_s$ has an $O(1)$ effect on $V_m$ of a passive cable if the dimensionless angular spatial frequency of $V_e$ ($\Omega = 2\pi f_s \lambda$) is larger than 1 and $1/L$, where $L$ is the cable's physical length normalized by its space constant $\lambda$. Both conditions are satisfied in these simulations because $L$ in our cable model is twice the cable space constant. This implies $1/L$ is one-half (less than one). In addition, the spatial profile of $V_e$ in Fig. 2D (although not exactly a sine wave) appears somewhat like a half-cycle of a periodic waveform. We can say, therefore, that $f_s \lambda \approx 1$ and thus $\Omega > 1$.

    If we move the stimulation site to different locations along the cable, the spatial angular frequency of $V_e$ appears to always be about one-half to one times the length of the cable (Fig. 3D). Thus the effect of $V_e$ on $V_m$ remains roughly $O(1)$ in all cases, as shown in the responses of the test neuron in Fig. 3C. This generic feature of the $V_e$ spatial profile generated by the cable population is due to conservation of current and the geometry of the one-dimensional volume conductor. Depolarizing input current reduces $V_e$ near the input site and the return of current back to the extracellular space restores $V_e$ to zero or positive values at locations of the cable distant from the input site. These $V_e$ effects can also be seen in the population-averaged $V_m$ responses by comparing responses that include ephaptic coupling (thick lines in Fig. 3B) to those that do not include ephaptic coupling (thin lines).

    For the different input locations in Fig. 3, the spatial pattern of $V_e$ changes dramatically in the extracellular region beyond the ends of the cable (not shown). In these simulations, the distance from the ends of the cable to electric ground is two times the space constant (i.e. same as physical length of cable). For off-center inputs (blue and green), $V_e$ is non-zero beyond the ends of the cable as it decays linearly to 0 mV at electric ground. For inputs to the center of the cable (red), $V_e$ is 0 mV at the ends of the cable and remains zero at all spatial locations beyond the ends of the cable. This is reminiscent of a "closed field" configuration. The responses to off-center inputs have $V_e$ spatial profiles that extend broadly beyond the ends of the cables and would be classified as an "open field" configuration (Lorente de Nò 1947). We note that for this linear problem (passive cable with current input), a more complicated spatial pattern of inputs can be constructed by taking the appropriate sum of responses to "point current" inputs.

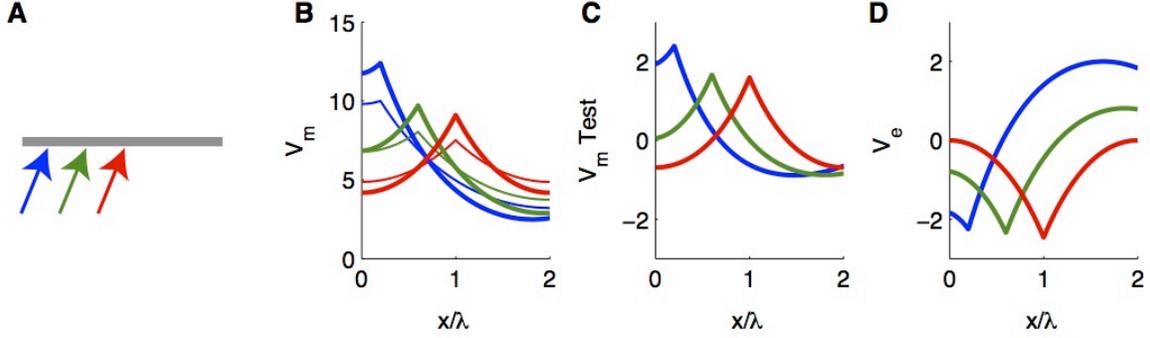

**Figure 3:** Passive cable response to constant input current for varying input location. **(A)** Input location is near end of cable (blue, same as Fig. 2), intermediate (green), or at center of cable (red). **(B)** Steady-state depolarization of population-averaged $V_m$. Thick lines are responses with ephaptic coupling ($\kappa = 1$) and thin lines are responses without ephaptic coupling ($\kappa = 0$). **(C)** Steady-state $V_m$ responses of "test neuron" that is coupled to population of cables via the extracellular voltage. **(D)** Steady-state $V_e$ responses produced by population of cables. Beyond the ends of the cable (not shown), $V_e$ decays linearly to electric ground (0 mV) for off-center inputs and remains at 0 mV for centered input.

*Ephaptic interactions are greatest for electrotonically compact cables*: It is known that the cable properties of a neuron model influence the amount to which its membrane potential is perturbed by a fixed, applied $V_e$. In particular, previous studies of passive cables embedded in $V_e$ found that the effect of $V_e$ on $V_m$ increases with the cable space constant (sinusoidal $V_e$ in Anastassiou et al. 2010; linear $V_e$ in Frölich and McCormick 2010). We show in Fig. 4 that the same result holds for endogenous $V_e$ generated in response to constant current input.

We keep the space constant of the cable population identical in all simulations (twice the physical length of the cables in the population), and obtain population-averaged $V_m$ and $V_e$ responses to constant, depolarizing current injection (Fig. 4A and Fig. 4C). This value of the space constant is our "reference" space constant $\lambda_{ref}$. The x-axis in all panels of Fig. 4 shows distance along the cables normalized by $\lambda_{ref}$.

Keeping the properties of the cable population unchanged, we vary the space constant $\tilde{\lambda}$ of the test neuron and observe changes in the test neuron's response to endogenous $V_e$. Specifically, we show that compact cables are more susceptible to ephaptic interactions (Fig. 4B). The value for the space constant used in previous simulations is half the physical length of the cable (denoted as $\lambda_{ref}$). If the space constant of the test neuron is reduced to half this value, i.e. an electrically longer cable, the ephaptic effect on $V_m$ decreases (green line). If the space constant is twice as large as the reference value, i.e. an electrically shorter cable (red line), the ephaptic effect on $V_m$ increases.

These changes of $V_m$ with space constant reflect that the test neuron's response to $V_e$ is a balance between local membrane currents that drive the membrane potential back to rest (0 mV in these figures since $V_m$ is plotted as deviation from rest) and axial intracellular currents that drive $V_i$ toward a constant spatial profile. In the limit of an electrotonically compact test neuron (large space constant), $V_i$ approaches a uniform spatial profile because intracellular current is easily redistributed along the cable. More precisely, for large $\tilde{\lambda}$ the deviation of $V_m$ from $E_{lk}$ for the test neuron approaches $-V_e + \frac{1}{L}\int_0^L V_e \, dX$. This value, which represents the upper bound on how much $V_e$ can perturb $V_m$ of the test neuron is shown as a thin black line in Fig. 4B.

In the opposite limit of an electrotonically long test neuron (small space constant), the local membrane currents dominate (relative to axial current flow). $V_m$ remains near its resting potential over most of the cable, but is perturbed near the input site and cable terminals. These observations regarding the sensitivity of the test neuron to $V_e$ also match our intuition from thinking of $V_e$ as a distributed "ephaptic current." Recall Eq. 2: an increase of $r_i$ (by reducing axial conductance, for instance) decreases the amplitude of the term $\frac{1}{r_i}\frac{\partial^2 V_e}{\partial x^2}$. Thus neurons with small space constants due to large internal resistance $r_i$ are unresponsive to $V_e$.

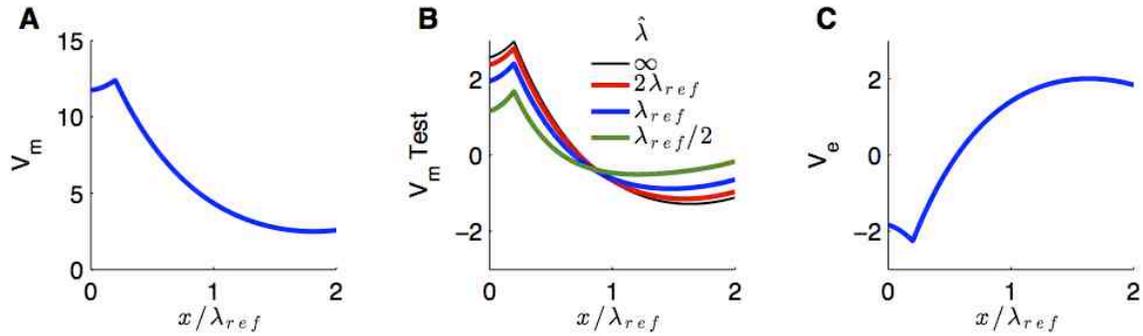

**Figure 4:** Passive cable response to constant input current for varying space constant $\hat{\lambda}$ of test neuron. **(A)** Steady-state depolarization of population-averaged $V_m$, same as Fig. 2B. **(B)** Steady-state response of "test neuron" $V_m$. Space constant $\hat{\lambda}$ of test neuron is varied so that it is smaller (green), same (blue) or larger (red) than space constant of cables in population. Space constant for cables in population is $\lambda_{ref}$ (half the physical length of the cable) in all simulations. **(C)** Steady-state $V_e$ response produced by population of cables, same as Fig. 2D. Abscissa shows dimensionless distance along cable relative to the reference space constant (half the physical length of all cables).

A novel feature of our model is that $V_e$ is not imposed as an exogenous input. $V_e$ is generated endogenously by the activity of a population of cables. We can also explore, therefore, how the spatial profile of $V_e$ depends on the space constant $\lambda$ of the cable population that generates $V_e$. We find the amplitude of $V_e$ increases with $\lambda$, as shown in Fig. 5C and Fig. 5F. Roughly speaking, we can imagine the cable population acts as a current dipole. A large space constant increases $V_e$ amplitude because it increases the dipole moment by allowing current to flow more easily within the cable (axially). As an illustration of this analogy, we mark the location of the current sink (site of current injection) by a black minus sign in Fig. 5C and Fig. 5F. We then computed the center of mass of source currents, where we defined source currents by $\max\left(0, -\frac{1}{r_i}\frac{\partial^2 V_e}{\partial x^2}\right)$. We marked these locations with colored plus signs. Consistent with analogy to the dipole moment, the location of the source moves away from the sink with increases in the cable space constant.

We vary $\lambda$ in two ways: by altering the membrane resistance $r_m$ (upper row of Fig. 5), or by altering the intracellular (axial) resistance $r_i$ (lower row of Fig. 5). The population-averaged $V_m$ responses change dramatically depending on which parameter is manipulated (compare the red curves in Fig. 5A and Fig. 5D, for instance, and note that the input current is identical in all simulations). In the top row, increases of $\lambda$ are associated with increases of $r_m$ and, consequently, increases of the cable input resistance (Rall 1977). In the bottom row, increases of $\lambda$ are associated with decreases of $r_i$ and, consequently, decreases of the cable input resistance. Recall that the coupling strength is inversely proportional to $r_i$, so for Fig. 5D-F we changed $\kappa$ to $1/4$ for the case of small $\lambda$ and $\kappa = 4$ for the case of large $\lambda$.

Although population-averaged $V_m$ responses change with these parameter manipulations, the $V_e$ responses remain similar for equal values of $\lambda$. Regardless of whether $r_m$ or $r_i$ is varied, the steady state distribution of transmembrane currents (steady-state, in these simulations) remains similar for a fixed $\lambda$. Since $V_e$ is generated by this current distribution, $V_e$ does not depend strongly on $r_m$ and $r_i$ as long as $\lambda$ remains at a fixed value.

In these simulations, the space constant of the test neuron is set to the reference value (half the physical length of the cable). Nonetheless, the perturbation of $V_m$ from rest increases for larger values of the cable population space constant (Fig. 5B and Fig. 5E). If we were to allow the test neuron space constant to co-vary with the cable population, we would see a "double effect" of ephaptic coupling. As the space constant of the population and test neurons increased together, the test neuron would be more susceptible to the effects of $V_e$ (recall Fig. 4B) and the amplitude of $V_e$ would increase (Fig. 5C and Fig. 5F).

**Figure 5:** Passive cable response to constant input current for varying space constant of the cable population. The test neuron space constant is fixed at the reference value (half the physical length of the cable, denoted $\lambda_{ref}$) in all simulations. Space constants of the cable population are varied by changing $r_m$ (top row) or $r_i$ (bottom row). **(A, D)** Steady-state depolarization of population-averaged $V_m$. **(B, E)** Steady-state $V_m$ response of test neuron. **(C, F)** Steady-state $V_e$ response produced by population of cables. The minus signs mark the location of the current sink (current injection). The plus signs mark the center of mass of source currents (see text for explanation). Abscissa shows dimensionless distance along cable relative to the reference space constant.

*Responses to sinusoidal current – Initial observations*: We have gained helpful initial insights by studying steady-state responses to constant current injection, but ultimately we are interested in dynamic ephaptic interactions (e.g. responses to trains of synaptic events). As a next step, therefore, we investigate responses of the cable population to sinusoidal current injection (Fig. 6). $V_m$, $\bar{V}_m$, and $V_e$ in this case are solutions to the boundary value problem Eq. 11-13. We visualize these solutions by plotting amplitude (Fig. 6A-C) and phase (Fig. 6D-F) as functions of normalized distance along the cable. Phase is in units of radians with zero-phase equal to the starting phase of the sinusoidal stimulus. We plot time courses of $V_m$, $\bar{V}_m$, and $V_e$ selected from three locations along the cable in Fig. 6G-I.

Responses to time-varying stimuli depend on the time constant $\tau = c_m r_m$ of the cable. In these simulations we set $\omega\tau = 0.1$, where $\omega$ is the stimulus frequency. This is an example in which the stimulus frequency is slow relative to the cable time constant. MSO neurons have exceptionally fast membranes (submillisecond time constants). We expect, therefore, that these simulations can provide intuition for MSO responses to stimulus frequencies as high as several hundred Hertz.

In the absence of ephaptic effects, the population-averaged $V_m$ attenuates along the length of the cable ($\kappa = 0$, black line in Fig. 6A). The speed at which $V_m$ propagates along the cable is evident in the roughly linear decay of phase along the cable (Fig. 6D). If ephaptic coupling is included ($\kappa = 1$, blue lines), the test neuron membrane potential and the extracellular voltage are non-zero. We remarked previously that ephaptic coupling effectively decreases the space constant of the cable (Rall 1977). This can be seen in the steeper attenuation of $V_m$ amplitude in for $\kappa = 1$. A smaller space constant is also associated with slower propagation of voltage along a cable (Koch 1998). Slowing of voltage spread due to ephaptic coupling is apparent in the steeper slope of the phase profile in Fig. 6E for the simulation with $\kappa = 1$ compared to the simulation with $\kappa = 0$.

The $V_e$ amplitude profile has two peaks (Fig. 6C) that correspond to two (roughly) anti-phase oscillations (note the abrupt, half-cycle phase transition in Fig. 6F). These anti-phase oscillations can be seen in time courses of $V_e$ by comparing responses near the input site to responses distant from the input site (Fig. 6I). This response profile is an indication that the cable population acts like a collection of synchronized current dipoles (Mc Laughlin et al 2010). The spatial location of the minimum of the $V_e$ amplitude profile is similar to the location of the half-cycle phase transition. For very low frequencies it would align with the location of the zero-crossing of the steady-state $V_e$ response in Fig. 2D.

The $\bar{V}_m$ response of the test neuron (Fig. 6B, E, H) has similar characteristics as the $V_e$ response. In particular, it is comprised primarily by two anti-phase oscillations (note the two peaks in the amplitude profile and the corresponding half-cycle phase transition). It can be helpful to distinguish these two "modes" by which ephaptic coupling drives $\bar{V}_m$ in the test neuron. On the proximal side, the test neuron membrane potential oscillates nearly in phase with the population-average membrane potential (compare green time courses in Fig. 6G and H). In contrast, the central and more distant regions of the test neuron oscillate anti-phase relative to the left side of the test neuron (compare red and cyan lines to green in Fig. 6H). In the transition region between these two oscillatory "modes," there is a minimum in the $V_m$ amplitude profile. For very low frequencies, this minimum would approach 0 mV at the location of the zero-crossing in the stationary response (Fig. 2C). In response to time-varying inputs, however, there is spread of voltage along the cable and $V_m$ is not equal to 0 mV at one fixed location for all time.

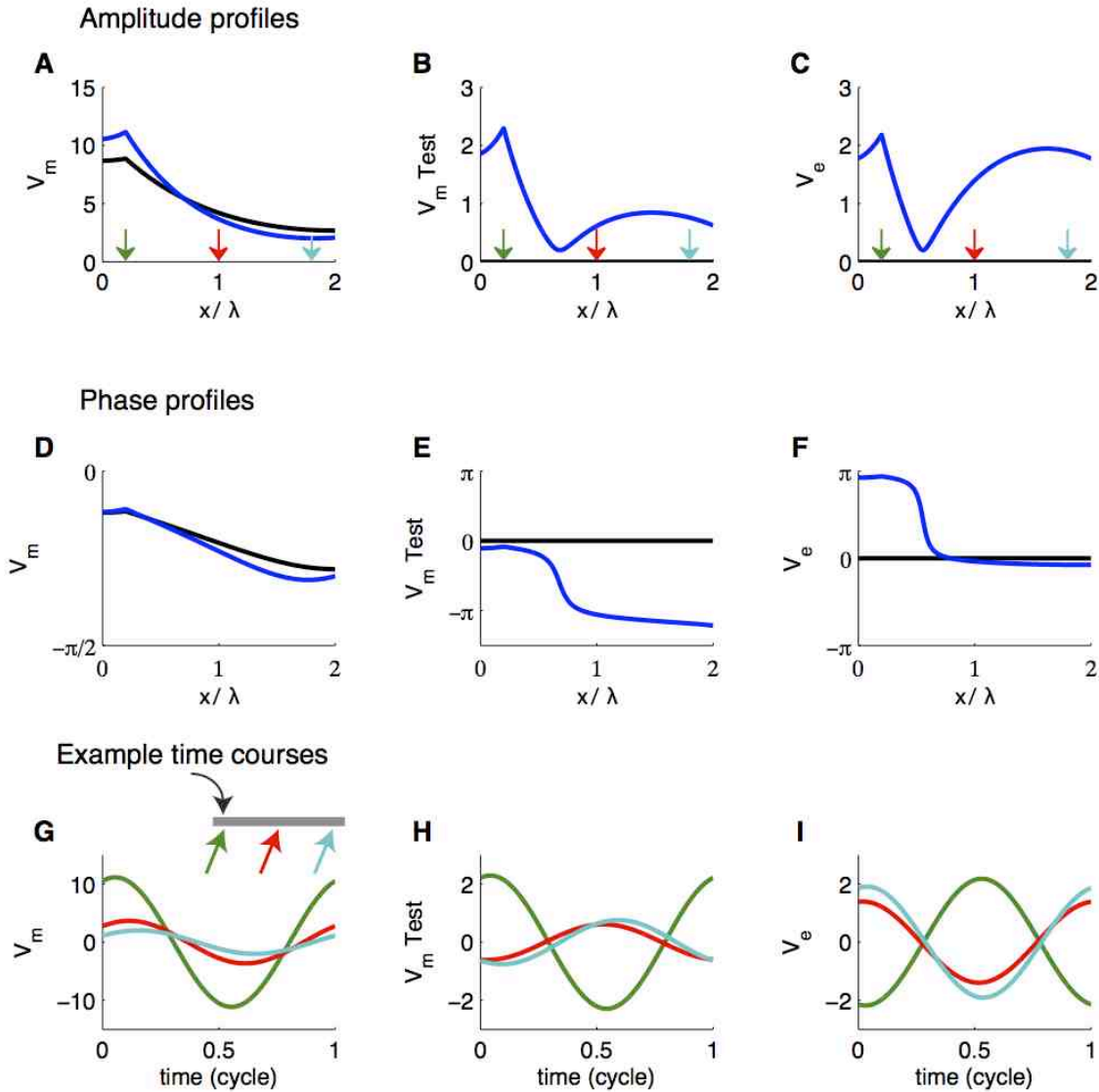

**Figure 6:** Passive cable response to sine-wave input current. Input frequency in dimensionless units is $\omega\tau$ = 0.1. **(A-C)** Amplitude profiles of cable population ($V_m$), test neuron ($V_m$), and extracellular voltage response ($V_e$). Membrane potentials are plotted as deviation from rest (units: mV). **(D-F)** Phase profiles cable population ($V_m$), test neuron ($V_m$), and extracellular voltage responses ($V_e$). Phase is in units of radians and zero-phase is referenced to the stimulus phase. In Panels A-F, the x-axis is the dimensionless distance along the cable (distance normalized by the space constant). Black lines show responses without ephaptic coupling ($\kappa$ = 0) and blue lines show responses with ephaptic coupling ($\kappa$ = 1). **(G-I)** Time courses of $V_m$, $V_m$, and $V_e$ plotted at three locations along the cable. Abscissa is one cycle of oscillations, ordinate is in units of millivolt (deviation from rest for population-averaged and test neuron). Schematic in G shows input location (black arrow). Time course locations are marked by colored arrows in A-C and G. They are X = 0.2, 1, and 1.8, where X = x/$\lambda$ is a dimensionless measure of distance along the cable.

*Attenuation of high frequency responses due to low pass cable dynamics:* Responses to higher frequency stimuli are attenuated by capacitative filtering of the passive cable (Fig. 7). The amplitude of population-averaged $V_m$ decreases with increasing frequency at all locations along the cable (Figs. 7A and 7D).

Extracellular voltage (Figs. 7C and 7F) and the test neuron $V_m$ (Figs. 7B and 7E) responses exhibit slightly more complex changes with stimulus frequency. We remarked previously that the $V_e$ responses to low frequency inputs are dipole-like and that the ephaptic interaction in these cases evokes two anti-phase "modes" of oscillation in test neuron $V_m$ of the test neuron. As stimulus frequency increases, these dipole-like response features are distorted. In particular, when the time scale of the stimulus and the test neuron's cable dynamics are similar (green curve, $\omega\tau =$ 1), then the two "modes" of oscillation interact via spread of membrane potential along the cable. We have provided user-friendly simulation code to the ModelDB website so that the interested reader can view movies of these time-varying solutions.

We remark that at specific positions along the cable (say, the point aligned with the minimum of the black curve in Fig. 7C), $V_e$ amplitude has a non-monotonic dependence on stimulus frequency. For the three locations we plot in Fig. 7F, however, $V_e$ amplitude attenuates monotonically with stimulus frequency.

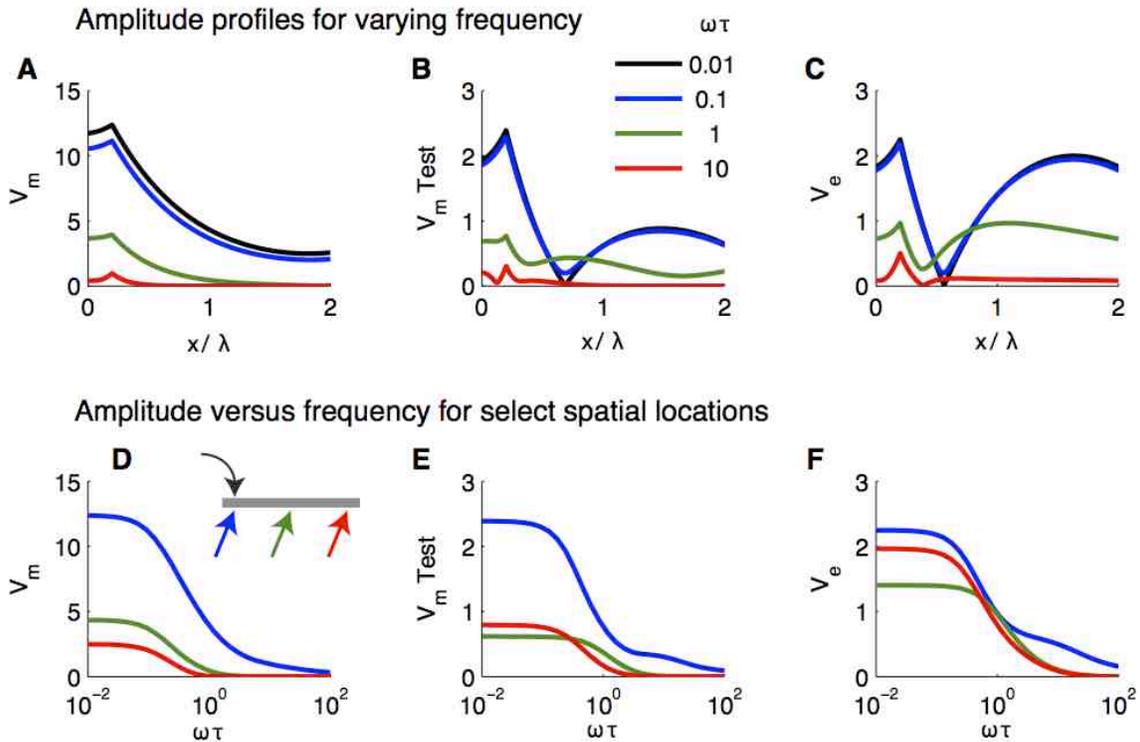

**Figure 7:** Attenuation of high frequency responses. **(A-C)** Amplitude profiles of cable population ($V_m$), test neuron ($V_m$), and extracellular voltage responses ($V_e$). Membrane potentials are plotted as deviation from rest (units: mV). Different lines represent responses to different stimulus frequencies (see legend in B). **(D-F)** Amplitudes of $V_m$, $V_m$, and $V_e$ responses plotted against stimulus frequency for three locations along the cable. Input location (black arrows) and response locations (colored arrows) are indicated by the schematic in D, they are X = 0.2, 1, and 1.8.

**Ephaptic interactions in a model of medial superior olive**

Initial simulations using a passive cable model have provided a basic understanding of the spatial and temporal patterning of ephaptic interactions in dendrite bundles. We next investigate ephaptic interactions in a biophysically-based model of the medial superior olive (MSO) to determine possible effects of $V_e$ coupling in a specialized nucleus in the auditory brainstem.

*Remarks on coupling parameter $\kappa$*: Recall that the coupling parameter depends on the ratio of extracellular to intracellular resistivity ($\rho$) and the packing density of neurons $\delta$ according to the relationship $\kappa = \rho\delta/(1-\delta)$. Extracellular resistivity $R_e$ in the auditory brainstem has not been measured, to our knowledge. As noted above, typical values of $\rho$ in models of local field potentials in cortex are often in the range of 2.2 to 3.5 (Holt and Koch 1999; Lindén et al. 2011; Reimann et al. 2013). It is plausible that $\rho$ in the MSO is larger due to the dense packing of myelinated fibers passing through the auditory brainstem, but we will use $\rho = 3$ as a reasonable estimate of the resistivity ratio.

We can estimate the packing density of neurons from an anatomical study of the MSO neurons in the gerbil (Rautenberg et al 2009). In that study, the mean soma diameter was 13 μm and the density of MSO neurons in a mature MSO slice was 7 cells per 100 μm. Consider, then, an idealized cross-section of MSO containing one column of 7 MSO somata (i.e. 100 μm length by 13 μm width). The packing density in this column is 0.715. In our simulations, we take $\delta = 0.7$ so that $\kappa = 7$ around the soma. Rautenberg et al. reported the diameter of dendrites in mature MSO slices was ~3 μm. The packing density of dendrites, and consequently the value of $\kappa$ in regions surrounding dendrites, is smaller than the values estimated above for somata. If we again consider 7 cells distributed in a 100 μm by 13 μm column, then the packing density of dendrites is $\delta = 0.038$. We set $\rho = 3$ (the same value as we used for the soma) and estimate the coupling strength for dendrites to be $\kappa = 0.12$. These estimated values of $\kappa$ near the soma and dendrites are marked in Fig. 1.

*Responses to monolateral synaptic excitation:* MSO neurons receive excitatory inputs that predominantly target their dendrites (Couchman et al. 2012). We begin our investigation of ephaptic coupling in MSO, therefore, by examining responses to a single excitatory synaptic event on one dendrite. The synaptic input depolarizes $V_m$ in the dendrites of the population of MSO neurons (~30 mV near the input site). $V_m$ amplitude attenuates as it spreads along the neuron (Fig. 8D, and evident in time courses at three locations along the neuron in Fig. 8A). $V_m$ in the soma is ~9 mV and the finite propagation speed is evident as the peak of $V_m$ is increasingly delayed as the post-synaptic potential propagates through the neuron.

The $V_e$ response(Fig. 8C) is negative near the input site (blue curve) due to the local transfer of positive ions from the extracellular domain into the intracellular domain (current sink). Near the soma and opposite dendrite, $V_e$ is positive (green and red curves) due to combined contributions of return currents distributed across the neuron (current sources). Recall that the coupling parameter in the dendrites is small ($\kappa = 0.12$). Nonetheless, the stimulation strength used in this simulation (maximal conductance is 27 mS/cm$^2$) suffices to generate $V_e$ amplitudes of $\pm$ 0.8 mV in the extracellular domain surrounding the dendrites.

The "test" MSO neuron receives no direct synaptic input; its membrane potential $V_m$ (Fig. 8B) is perturbed by the spatio-temporal pattern of the surrounding extracellular voltage. $V_m$ increases near the stimulation site and decreases in the soma and distal dendrite. The peak ephaptic "excitation" is ~1 mV. This illustrates that the millivolt-scale $V_e$ responses observed *in vivo* in the MSO (Mc Laughlin et al. 2010, e.g.) could, in principle, represent a non-synaptic mechanism by which MSO neurons could induce millivolt-scale perturbations in membrane potential of neighboring neurons.

These simulations illustrate the dynamics of $V_m$, $V_m$, and $V_e$ responses to simulated synaptic inputs. Many of the main qualitative features, however, were already present in the steady state passive cable simulations presented at the outset. To highlight the useful insights provided by the steady-state cable model, we plot spatial profiles of the maximum deviation from rest for the MSO neuron model in the bottom row of Fig. 8. These results can be compared to steady state passive cable responses (Fig. 2). Simulations without ephaptic coupling are shown in black and simulations with ephaptic coupling are shown in blue.

The deviation from resting voltage of the MSO neuron and passive cable models share many of the same qualitative features. For instance, in both cases the test neuron $V_m$ amplitude in the soma (or center of passive cable) is negative. This indicates that ephaptic coupling (in this scenario of monolateral dendritic excitation) diminishes the response at the soma and may raise the threshold for MSO spiking. We will explore this in more detail below (see Figs. 10 and 11). There are, of course, quantitative differences. The profiles of $V_m$, for example, show that ephaptic effects are smaller in the MSO model than in the cable model (compare blue and black lines in Fig. 8D and Fig. 2B). The weaker ephaptic effect in the MSO model is likely due to the small coupling coefficient $\kappa$ for MSO dendrites. Perhaps surprisingly, the $V_m$ responses of the test neuron is largest in the dendrite (near the site of synaptic input) despite the small $\kappa$ value there.

## Example time courses

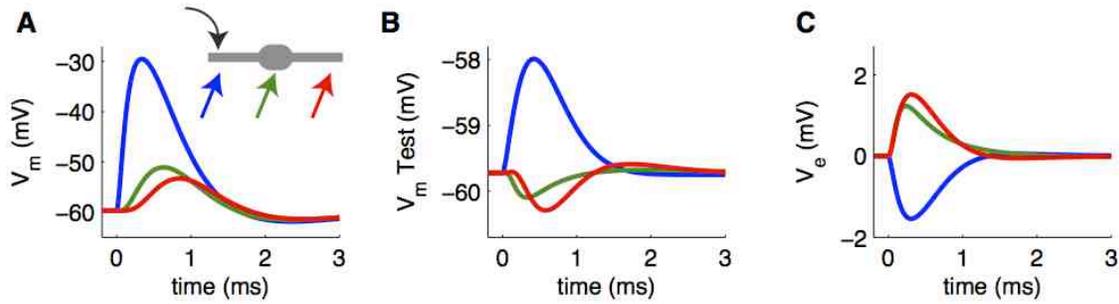

## Maximum deviation from rest

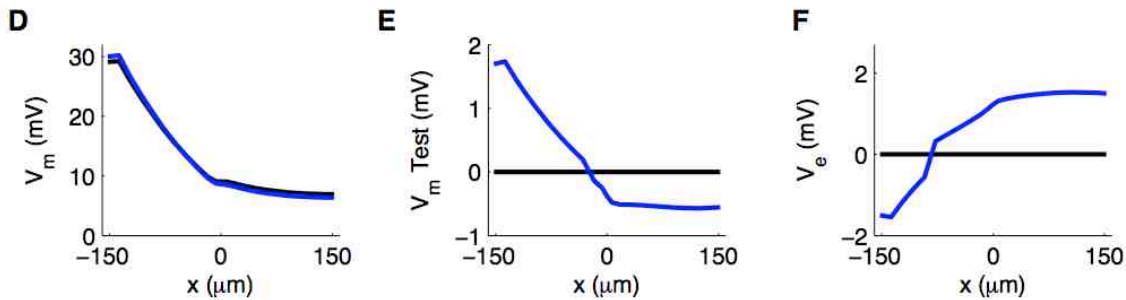

**Figure 8:** MSO response to monolateral synaptic excitation. **(A-C)** Example time courses of $V_m$ in MSO population, $V_m$ in the test neuron, and extracellular $V_e$ at three locations along the neuron. Schematic in A shows these locations (left dendrite near the input site, soma, and right dendrite) and input location (left dendrite). Input to MSO population is a simulated excitatory synaptic event located 137.5 μm from the center of the soma with peak conductance 27 mS/cm$^2$. Test neuron receives no synaptic stimulation. **(D-F)** Spatial profiles of maximum deviation from resting voltage for $V_m$, $V_m$, and $V_e$. in response to monolateral synaptic excitation. Ordinate is deviation from resting voltage and can be compared to steady state passive cable responses in Fig. 2. Results for simulations without ephaptic coupling are in black, results for simulations that include ephaptic coupling are in blue.

*Responses to bilateral synaptic excitation:* In natural listening conditions, MSO neurons receive excitation on both dendrites (from sounds arriving in both ears). In Fig. 9 we show responses to coincident bilateral inputs (simulated excitatory synaptic events that arrive simultaneously on both dendrites). Example voltage time courses are shown in the top row and the bottom row shows the maximal deviation from resting voltage.

The synaptic inputs depolarize the $V_m$ by ~30 mV near the synaptic site and a summed depolarization of ~15 mV in the soma (Fig. 9A, D). Responses to bilateral inputs differ slightly from the linear superposition of monolateral responses due to the presence of voltage-gated low threshold K current in the dendrites and soma.

A striking difference in these simulations compared to responses to monolateral inputs shown in Fig. 8 above is that $V_e$ is spatially localized (Fig. 9C, F). $V_e$ reaches a maximum value ~ 2.5 mV near the soma, but near the distal reaches of the dendrites $V_e$ decreases to 0 mV. The symmetric arrangement of membrane currents produces a "closed-field" with no volume conduction beyond the dendrites' terminal ends. Given the small $V_e$ and small coupling parameter surrounding the dendrite, ephaptic effects on the dendrites might not be expected. This is not the case. The depolarization of $V_m$ in the dendrites of the test neuron is, in fact, roughly twice as large (in amplitude) as the hyperpolarization in the test neuron soma (compare blue curve to green curve in Fig. 9B).

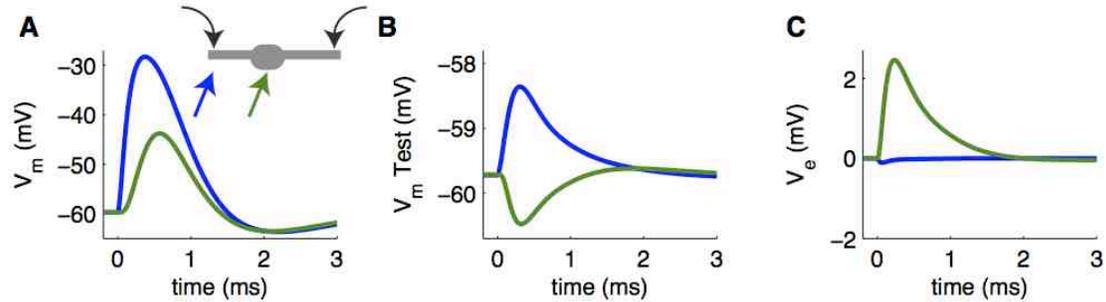

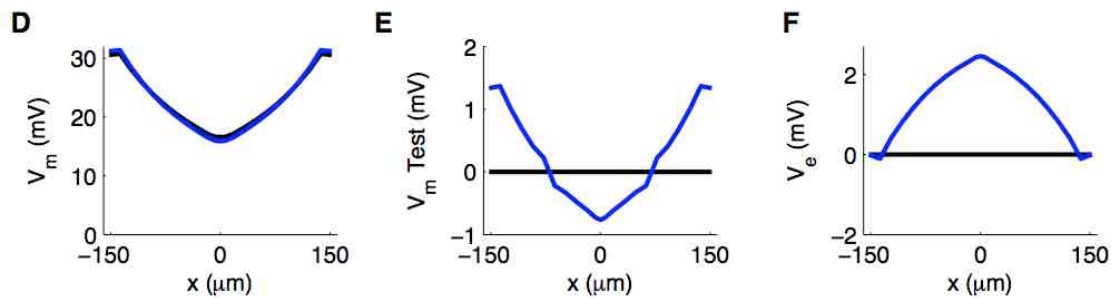

**Figure 9:** MSO response to bilateral synaptic excitation. **(A-C)** Example time courses of $V_m$ in the MSO population, $V_m$ in the test neuron, and extracellular $V_e$ at two locations along the neuron. Schematic in A shows these locations (left dendrite near the input site, soma) and input locations (both dendrites). Responses on right dendrite are identical to those on the left dendrite and are not shown. Inputs to the MSO population are two simulated excitatory synaptic events located 137.5 μm from the center of the soma on both dendrites. Synaptic events have identical maximal conductance (27 mS/cm$^2$) and onset time (0 ms). **(D-F)** Spatial profiles of maximum deviation from resting voltage for $V_m$, $V_m$, and $V_e$ in response to bilateral synaptic excitation. Results for simulations without ephaptic coupling are in black, results for simulations that include ephaptic coupling are in blue.

*Ephaptic coupling influences MSO spike initiation:* We have shown that endogenously-generated $V_e$ can perturb the membrane potential of a "test" MSO neuron embedded in the extracellular bath. Does this ephaptic interaction suffice to alter spiking outputs of MSO neurons? By adding a spike-initiation zone (SIZ) to the test neuron model (see Methods), we can investigate how SIZ membrane potential and spiking activity is influence by ephaptic interactions.

We created a situation in which $V_e$ is generated from MSO neuron models that do not include an SIZ. We then embedded a test neuron with a SIZ into this $V_e$ field (see Methods). The test neuron (including its SIZ) feels the influence of $V_e$ but does not feedback and contribute to it. We are treating this one neuron's contribution to $V_e$ as sufficiently small that it can be ignored.

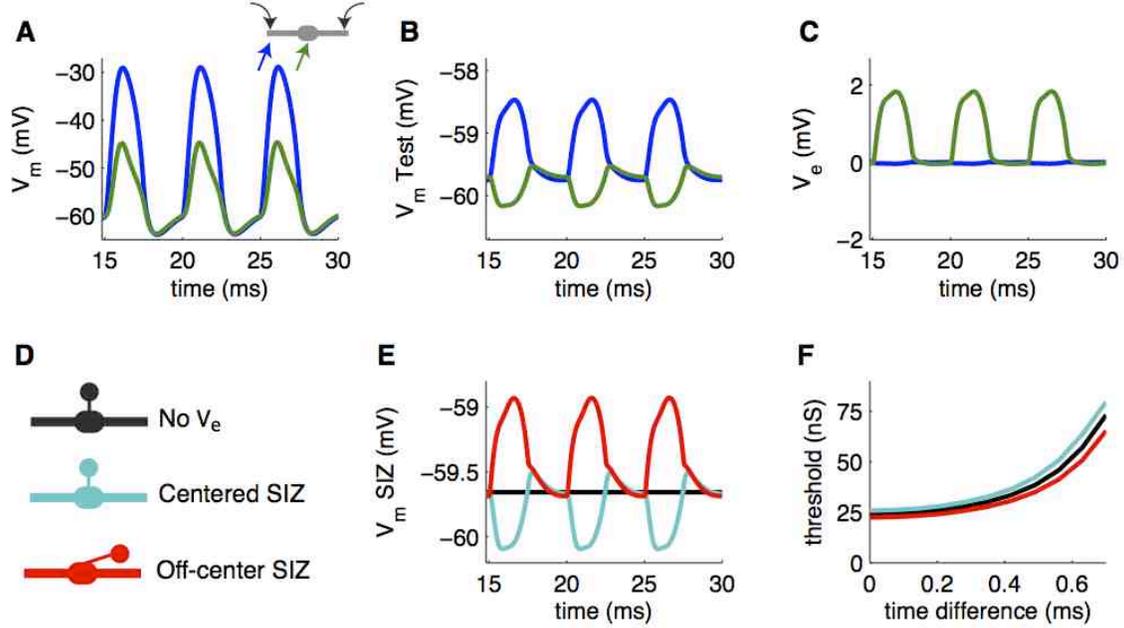

**Figure 10:** Response of MSO to periodic bilateral excitation. Test neuron includes a spike initiation zone (SIZ). **(A-C)** Time courses of $V_m$ in the MSO population, $V_m$ in the test neuron, and extracellular $V_e$ at two locations along neuron. Schematic in A shows these locations: left dendrite near the input site (blue) and soma (green). Responses on right dendrite are identical to those on the left dendrite and are not shown. Waveform of excitatory conductance to MSO population is 200 Hz rectified sine wave (identical on both dendrites, located 137.5 μm from the center of the soma). Maximal conductance is 20 mS/cm$^2$). Response is shown after 15 ms to avoid onset effects. **(D)** Schematic diagram illustrating the different SIZ and $V_e$ configurations tested: no ephaptic coupling (black), ephaptic coupling and SIZ located in alignment with the soma at x = 0 μm (cyan), ephaptic coupling and SIZ located ~100 μm away from soma (red). **(E)** Membrane potential response in SIZ for three different model configurations. $V_e$ coupling increases $V_m$ in the SIZ for off-center SIZ and decreases $V_m$ for centered SIZ. **(F)** Conductance threshold of test neuron in response to bilateral excitation. Inputs to test neuron are a pair of synaptic events (alpha function with 0.2 ms time constant, see Eq. 20) with time difference varied (x-axis).

We stimulate the MSO population with identical, synchronized excitatory conductance on both dendrites. The $V_m$ response is a large ~30 mV depolarization in the dendrites that attenuates to a ~15 mV depolarization in the soma (Fig. 10A). The input conductance is a 200 Hz rectified sine wave identical on both dendrites (peak amplitude is 20 mS / cm$^2$). We use the rectified sine waveform as a simplified, population-averaged representation of excitatory drive to MSO in response to a pure tone stimulus.

The MSO population generates a periodic $V_e$ response with ~1.8 mV positive going oscillations around the soma and much smaller negative-going oscillations around the distal ends of the dendrites (Fig. 10C). In response to this endogenous $V_e$ field, the test neuron exhibits ~1.5 mV positive-going oscillations in $V_m$ in the dendrites and smaller ~0.5 mV negative-going oscillations in the soma (Fig. 10B). Note that the test neuron does not receive any synaptic input in these simulations so these $V_m$ changes are strictly due to ephaptic coupling.

The test neuron includes a SIZ in these simulations and $V_e$ perturbs the SIZ's membrane potential (Fig 10E). In these and subsequent simulations, we compare three SIZ configurations illustrated in Fig. 10D: the "control" condition of no ephaptic coupling (black), ephaptic coupling and SIZ aligned with the soma (red), and ephaptic coupling and SIZ aligned with a position ~100 μm from the soma (cyan).  If the SIZ is located near the soma, its membrane potential exhibits negative-going oscillations similar to $V_m$ in the soma of the test neuron (compare cyan line in Fig. 10E to green line in Fig. 10B).  If the SIZ is located away from the soma, its membrane potential exhibits positive-going oscillations similar to the response of the test neuron dendrite (compare red line in Fig. 10E and blue line in Fig. 10B).  Note that in both cases the SIZ is connected via the same internal, axial resistance to the soma of the MSO neuron model; the SIZ position only determines the $V_e$ that the SIZ "sees."

These simulations reveal that $V_e$ can increase or decrease the SIZ membrane potential depending on the position of the SIZ.  Do ephaptic interactions alter spiking activity in MSO?  We used conductance threshold as a measure of neuron excitability and found that ephaptic effects modulate the threshold curve depending on the location of the SIZ (Fig. 10F).  If the SIZ is near the soma, we saw in Fig. 10E that $V_e$ hyperpolarizes the SIZ membrane potential.  This translates to an increase in threshold (diminished excitability) for all time differences tested.  In contrast, the depolarizing effect of $V_e$ on the SIZ located near the distal dendrites translates to a decrease in threshold, i.e. enhanced excitability.

The test neuron in these simulations received synaptic excitation in the form of excitatory (alpha function) synaptic events arriving on the two dendrites 15 ms after the start of the periodic input to the MSO population.  The 15 ms delay ensured that any transient onset dynamics are avoided.  We varied the difference in the timing of the two synaptic inputs to the test neuron (*x*-axis).  A time difference of 0 ms represents coincident bilateral inputs to the test neuron.  In this case the synaptic event times in the test neuron match the onset of one cycle of the rectified sine wave conductance input to the MSO population.  Time differences larger than 0 ms (positive values on the *x*-axis of Fig. 10F) represent bilateral inputs to the test neuron that are not coincident.  The synaptic event on one dendrite arrives earlier than the onset of the rectified sine wave input to the MSO population and the other synaptic event trails the onset of the rectified sine wave.

Conductance threshold is the smallest peak conductance needed to generate a spike in the SIZ.  In the absence of ephaptic effects, thresholds increase with sub-millisecond increases in synaptic time difference (black line in Fig. 10F).  This is confirmation that the model neuron, like MSO neurons, act as a coincidence detector.  For the time differences tested in these simulations, ephaptic interactions decrease spike threshold by approximately ~10% for off-center SIZ and increase threshold by ~10% for centered SIZ.

*Ephaptic coupling can entrain MSO spike timing*: The small changes in spike threshold measured above can result in changes in spike timing.  Specifically, we observed that periodic $\underline{V_e}$ can entrain a spontaneously firing neuron (Fig. 11A).  In these simulations, we provided the test neuron with a random train of excitatory synaptic events that caused the test neuron to fire spontaneously at a rate of 20 spikes per second in the absence of ephaptic coupling (black line).  Firing rate is presented as the cycle histogram of the test neuron's response.   In the absence of ephaptic interactions, there is no 200 Hz "rhythm" to entrain spike times, thus the cycle histogram is flat for the spontaneously firing neuron.

When we repeated the simulation in the presence of a 200 Hz endogenously-generated $V_e$, we found that ephaptic interactions temporally modulate spike timing. Consistent with our previous results, the effect of $V_e$ differs depending on the location of the SIZ. For the centered SIZ (cyan line), ephaptic coupling reduces firing rates during the first half of the 5 ms cycle and increases firing rates during the second half of the cycle. The cumulative effect of this $V_e$-induced firing rate modulation is a decrease of ~1.4 spike per second. For the off-center SIZ, the effect is opposite (red line). Ephaptic coupling increases firing rates during the first half of the 5 ms period and decreases firing rates thereafter. The cumulative change is an extra ~1.7 spikes per second due to ephaptic coupling. Simulations that include ephaptic coupling and the off-center SIZ produce spikes that are more likely to occur in phase with the periodic input to the MSO population. Thus the ephaptic interaction can entrain the test neuron's spikes, even though inputs to the test neuron are random and do not have a periodic structure.

To construct the input to each dendrite of the test neuron, we combined 10 independent realizations of a homogeneous Poisson process (100 Hz event rate) to generate 10 independent input trains of "spike times." We then convolved these event times with an alpha function (0.2 ms time constant, see Eq. 20). Each "unitary" synaptic event had a peak conductance of 10 mS/cm$^2$ that increases the SIZ membrane potential by 4.5mV. The simultaneous arrival of four such events (two per dendrite) can evoke a spike in the SIZ as can the arrival of six such events on a single dendrite.

*Ephaptic coupling influences MSO coincidence detection*: As a final test of ephaptic interactions, we simulated the standard measure of MSO neuron's tuning to sound location: time difference tuning curves (Fig. 11B). Inputs to the test neuron were constructed from inhomogeneous Poisson processes with the Poisson rate given by a 200 Hz rectified sine wave. The inputs to the test neuron and MSO population have the same timing relative to one another, but there can be a time difference between the inputs on opposite dendrite. The *x*-axis reports the time difference in the bilateral inputs, a time difference of 0 ms indicates inputs that are identical on the two dendrites. A non-zero time difference indicates a phase difference between the sine wave rate functions.

The test neuron responds maximally when the Poisson rate functions are in phase on the two dendrites (0 ms time difference). This is another indication that the MSO model neuron acts as a coincidence detector with submillisecond temporal precision. In the absence of ephaptic coupling, the maximal firing rate is 166 Hz and decreases for larger time differences (black line in Fig. 11B). When ephaptic coupling is included, the effect on time difference tuning depends on the SIZ location in a manner consistent with our previous tests. Ephaptic coupling and the centered SIZ combine to have an "inhibitory" effect that reduces firing rates (cyan line, firing rate at 0 ms time difference is 159 spikes per second). Ephaptic coupling and the off-center SIZ combine to have an "excitatory" effect that increases firing rates (red line, firing rate at 0 ms time difference is 174 spikes per second). Taken together, the results in Fig. 11 illustrate that the nonlinear nature of spike generation can amplify small changes in membrane potential. Millivolt-scale ephaptic interactions can plausibly alter spike activity in neurons and circuits.

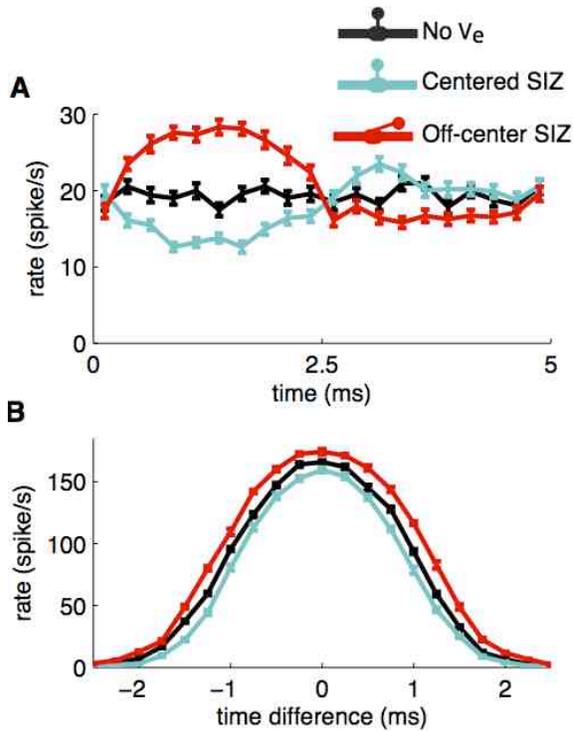

**Figure 11:** Ephaptic effects on spike timing and coincidence detection. **(A)** Cycle histogram of test neuron response to bilateral trains of excitatory synaptic events. Event times of input trains are generated from independent homogeneous Poisson processes (10 trains per dendrite. Event rate for the underlying Poisson process is 100 events per second. Average firing rates calculated in 0.25 ms bins from 50 simulations of responses to 10 second long input trains. Error bars are standard error of the mean. **(B)** Time difference tuning curves for test neuron's response to bilateral trains of excitatory synaptic events. Event times of input trains are generated from independent inhomogeneous Poisson processes (10 trains per dendrite). Event rate for the underlying Poisson process is a 200 Hz rectified sine wave with an average event rate of 100 events per second. The timing difference of the rectified sine wave Poisson rates is shown on the x-axis with 0 ms time difference representing rate functions to both dendrites that are in phase relative to one another and relative to the input to the MSO population. Firing rates were computed from responses to 10 second long inputs. Mean and standard error of the mean (error bars) were obtained by dividing these responses into 10 segments and counting spikes in each one second long subintervals. In all simulations (A and B): Unitary events synaptic events are alpha function conductances with 0.2 ms time constant and 10 mS/cm$^2$ maximal conductance, model configurations and color code are same as in Fig. 10D. Input to MSO population (generators of $V_e$) is 200Hz rectified sine wave excitation (20 mS/cm$^2$ maximal conductance) with same bilateral time differences as input to test neuron.

# DISCUSSION
## Summary of main findings
Neural activity generates transmembrane currents that, in turn, generate spatiotemporal patterns of extracellular voltage. Nearby neurons are embedded in this shared, endogenously-generated $V_e$. $V_e$ can provide, therefore, a channel for non-synaptic communication between neurons. Following the insightful early work of Arvanitaki (1942), we refer to this phenomenon as *ephaptic* coupling.

We have developed and analyzed a model of ephaptic interactions in an idealized one-dimensional setting (computer code available on ModelDB). We considered a population of identical passive cables as an idealized model of uniform, unbranched neurons receiving identical inputs. We illustrated how ephaptic coupling in this system can be described with an extension of cable theory that accounts for dynamical coupling between one-dimensional intra- and extracellular domains. We identified the strength of extracellular coupling by introducing a new dimensionless parameter $\kappa = \rho\delta/(1-\delta)$, where $\rho$ is the ratio of volume resistivities $R_e/R_i$ and $\delta$ is the packing density of neurons and offered plausible estimates for $\kappa$.

We found that the idealized neuron population could generate millivolt-scale $V_e$ and this $V_e$ could induce millivolt-scale perturbations in the membrane potential of an additional "test" neuron embedded in the endogenously-generated $V_e$. We varied the spatial profile of $V_e$ by varying the localized stimulus input site and found that the amplitude of the $V_m$ perturbation remained relatively constant (Fig. 3). We also found that electrotonically compact cables (large space constant $\lambda$) experience larger changes in their $V_m$ due to coupling to $V_e$ than electrotonically long cables (small space constant $\lambda$) (Fig. 4) and compact cables generate larger amplitude $V_e$ responses (Fig. 5). For sinusoidal inputs, ephaptic effects at low frequencies were similar to steady state responses (Fig. 6) and attenuated in response to higher frequency inputs due to the low-pass dynamics imposed by the passive cable (Fig. 7).

We applied the same one-dimensional idealization to model $V_e$ generated by simultaneously activated and spatially aligned neurons in the medial superior olive (MSO) of the auditory brainstem. These neurons have a relatively simple structure: a soma and two dendrites extending away from the soma with minimal branching. The qualitative features of ephaptic coupling in the passive cable model and this more biophysically-detailed model were broadly similar. Our simulations used physiologically plausible parameters and fast synaptic conductances and demonstrated that MSO neurons can generate millivolt-scale $V_e$ responses that induce millivolt-scale perturbations in a "test" MSO neuron (Figs. 8 and 9).

The MSO is a critical early stage of binaural processing and it can extract information regarding the location of sound sources in the environment. MSO neurons receive inputs arriving from both ears and they are sensitive to submillisecond timing differences in these inputs. We were particularly interested, therefore, to determine whether ephaptic effects and their instantaneous nature can alter the precise coincidence detection computation performed by MSO neurons. We tested ephaptic effects on MSO spiking by adding a spike initiation zone (SIZ) to a test neuron embedded in the endogenously-generated (simulated) $V_e$. We found that millivolt-scale ephaptic effects can change the spike output of MSO neurons. The location of the SIZ in the spatially-distributed $V_e$ is critical. We compared two configurations: a "centered" SIZ aligned with the soma and an "off-center" SIZ located ~100 μm away from the soma (see schematic in Fig. 10D) . $V_e$ can act in an "excitatory" or "inhibitory" manner depending on the SIZ position. In particular, for an SIZ located away from the soma, $V_e$ produced by in phase bilateral inputs to the MSO population depolarized $V_m$ in the SIZ (Fig. 10E), decreased thresholds for spike generation (Fig. 10F), entrained spike times to the ongoing $V_e$ oscillation (Fig. 11A), and increased the gain of a simulated time-difference tuning curve (Fig. 11B). If the SIZ is aligned with the soma, $V_e$ has opposite effects and suppresses spiking activity.

**Relation to previous work**

Our model relies on the idealization that neural dynamics in the presence of ephaptic interactions can be described by coupling a one-dimensional intracellular domain (cable core conductor) to a one-dimensional extracellular volume conductor. We arrived at this simplified geometry by taking a mean-field view of a large population of identical cables organized in parallel to one another and receiving identical inputs. This construction may appear overly simplistic, but it is inspired by pioneering studies of endogenously-generated $V_e$ (Rall and Shepherd 1968, Nicholson and Llinás 1971) and we have recently used a similar formulation to study *in vivo* $V_e$ in the auditory brainstem of cats (Goldwyn et al 2014). Alternative methods have been formulated that include more realistic neural morphologies and extracellular volume conductors (Malik 2011, Agudelo-Toro and Neef 2013), but these methods have the drawback of requiring substantial computing resources. In addition, the one-dimensional intracellular and extracellular domains facilitate visualization and analyses of simulation results. The validity of the one-dimensional volume conductor model could be evaluated in future studies with these more sophisticated computational methods.

A number of our observations concur with earlier studies of $V_e$ effects on passive cables. Rall (1977) pointed out that coupling to $V_e$ decreases the cable space constant from $\sqrt{r_m/r_i}$ to $\sqrt{r_m/(r_i + r_e)}$. Recent studies have pointed out that increasing the space constant of a cable (larger $r_m$ and/or smaller $r_i$) increases the effect of $V_e$ on $V_m$ (Anastassiou et al 2010, Frölich and McCormick 2010). We identified the upper bound of the ephaptic effect for steady state responses to constant localized input as $-V_e + \frac{1}{L}\int_0^L V_e \, dX$ (see Fig. 4B).

We showed that a periodic $V_e$ can alter spike timing in MSO model neurons (Fig. 11A). Recent studies have reported similar results *in vitro* and in simulation studies (Frölich and McCormick 2010, Anastassiou et al 2011). Specifically, these studies applied periodic extracellular fields to cortical slice preparations and demonstrated that spike timing can be entrained to the "rhythm" of ongoing, periodic $V_e$. A novel finding in our study is that the location of the SIZ matters when considering the effect of $V_e$ on spike timing. $V_e$ can indeed entrain neurons and enhance spike synchrony (Fig. 11A, off-center SIZ). It is also possible, however, that the SIZ could be located in a region of extracellular space where $V_e$ decreases the SIZ membrane potential and suppresses spike output (Fig. 11A, centered SIZ). This could reduce spike time synchrony across a population. We did not consider the contribution of spikes to $V_e$ because they are not prominent in MSO field potentials. A recent modeling study of layer V pyramidal cells has demonstrated that $V_e$ generated by spikes is capable of altering spike timing in simulations (Stacey et al. 2015).

**Implications for functional ephaptic coupling**

Applied extracellular voltage affects cell-level and circuit-level neural activity; this is the principle by which neural prostheses such as cochlear implants, retinal implants, and deep brain stimulation operate. We are concerned with a complementary question: does endogenously-generated $V_e$ act as a form of non-synaptic, global coupling to alter neural dynamics?

Studies in diverse neural structures have identified ephaptic coupling as a means of fast, non-synaptic inhibition (teleost Mauthner-cell system in goldfish, Weiss et al. 2008; olfactory receptor neurons in insects, Su et al. 2012; pinceau structure surrounding axon initial segment of cerebellar Purkinje cell, Blot and Barbour 2014). Figure 2 illustrates how ephaptic coupling could operate in an inhibitory manner in a bundle of neurons. Excitatory inputs arriving near the terminal of a neural structure (a distal dendrite site, e.g.) hyperpolarize the opposite end (e.g., soma and initial segment) of neighboring neural structures through the ephaptic interaction.

Figure 2 also illustrates that ephaptic interactions could promote local excitation. Excitatory inputs arriving near the end, say proximal end, of a neural structure locally depolarize the proximal end of nearby neurons through the ephaptic interaction. This local depolarization could be enhanced further in dendritic bundles with regenerative currents and provide a non-synaptic mechanism for simultaneously exciting dendrites in a local population. Bokil et al. (2001) demonstrated this type of "excitatory" ephaptic coupling in simulations of a bundle of axons. They found, using a mean field formulation similar to ours, that spikes in an axon can promote spike generation and spike time synchrony in nearby axons. The capacity for $V_e$ to depolarize nearby neurons via an "excitatory" ephaptic effect has also been proposed as a mechanism for epileptogenesis in the hippocampus (Traub et al. 1985, Zhang et al. 2014).

Our MSO simulations that included a spike initiation zone (SIZ) in the test neuron demonstrated that even small ephaptic effects (millivolt-scale) can alter spiking activity (Fig. 11). The nonlinear nature of spike generation can amplify ephaptic coupling (see also: Radman et al 2007, Anastassiou et al. 2011). We found that the anatomy of the spike generator matters for neurons embedded in a spatially-varying $V_e$. Ephaptic coupling can have opposing "excitatory" or "inhibitory" effects depending on the orientation of axons and the site of spike generation. We are not aware of a comprehensive study of axon anatomy in the MSO. This makes it difficult for us to make specific predictions regarding possible functional effects of ephaptic coupling in MSO. A recent study has examined spike generation in MSO axons and the anatomy of the MSO axon initial segment (Lehnert et al 2014). Further work in this direction would contribute to our understanding of ephaptic coupling the MSO. Auditory brainstem neurons in the chick exhibit plasticity in the length of the axon initial segment (specifically, the distribution of Na channels) (Kuba et al 2010). This raises the possibility that ephaptic effects could be modulated over time.

Our cable theory-based study of ephaptic interactions among passive cables has informed our understanding of ephaptic interactions in MSO neurons. We view the MSO as a useful "model system" to explore these effects because $V_e$ is large and sound-evoked *in vivo*, spatio-temporal features of the $V_e$ response can be modeled with the mean-field approximation and a one-dimensional volume conductor (Goldwyn et al 2014), and MSO neurons perform a known computation. That computation – coincidence detection of binaural inputs – is typically studied by recording (or simulating) time-difference tuning curves analogous to Fig. 11B. We found that ephaptic coupling could increases or decrease the gain of the time-difference tuning curve, depending on the location of the spike generator (but the tuning curve width was not changed appreciably). Our simulations establish a "proof of principle" that ephaptic interactions can alter binaural processing in the MSO. We cannot yet identify a functional role for ephaptic coupling in the MSO, but this remains an intriguing avenue for future research.

## Acknowledgements

Large-scale simulations were supported by computing resources at the New York University High Performance Computing Center and an allocation of computing time from the Ohio Supercomputer Center.

## Grants

This research has been supported by funding from the National Institute on Deafness and Other Communication Disorders: F32 DC012978 (JHG) and R01 DC008543 (JR).

## Disclosures

The authors declare that they have no conflict of interest, financial or otherwise, pertaining to this work.